\documentclass{jfm}

\usepackage{graphicx}
\usepackage{newtxtext}
\usepackage{newtxmath}
\usepackage{amsmath}
\usepackage{natbib}
\usepackage{xcolor}
\usepackage{hyperref}
\usepackage{placeins}
\usepackage[thinc]{esdiff}
\usepackage{booktabs}
\usepackage{makecell}
\hypersetup{
    colorlinks = true,
    urlcolor   = blue,
    citecolor  = blue,
}

\newcommand{\RomanNumeralCaps}[1]
\nolinenumbers

\title{Gravity-driven viscous flow over partially lubricated bed}

\author{Joshua H. Rines\aff{1,2}
  \corresp{\email{jrines@stanford.edu}},
  Ching-Yao Lai\aff{1,2},
  Yongji Wang\aff{1,2,3}}

\affiliation{\aff{1}Department of Geophysics, Stanford University, Stanford, CA
\aff{2}Department of Geosciences, Princeton University, Princeton, NJ
\aff{3}Department of Mathematics, New York University, New York, NY}

\begin{document}
\maketitle

%%%%%%%%%%%%%%%%%%%%%%%%%%%%%%%%%%%%%%%%%%%%%%%%%%%%%%%%
%%%%%%%%%%%%%%%%%%%%%%% ABSTRACT %%%%%%%%%%%%%%%%%%%%%%%
%%%%%%%%%%%%%%%%%%%%%%%%%%%%%%%%%%%%%%%%%%%%%%%%%%%%%%%%
\begin{abstract}
We present an investigation into the response of a viscous fluid flowing over a sloped bed across a spatially finite patch of basal lubrication.  We present a simple analytical model that captures the fundamental structure of such lubrication-induced stress and velocity perturbations in Newtonian fluids, as well as scaling arguments and numerical experiments that extend our analysis to power-law fluids.  These analyses concisely reveal the underlying relationships between the system parameters (fluid thickness, $h$, slope, $\alpha$, slippery patch length, $\ell$, and sliding condition outside of the slippery patch, $\gamma$) and the magnitude and spatial extent of the resulting perturbed stresses, $\tau_{xx}$, and velocities, $u_p$.  From these results, we conclude that the induced stresses are exponentially decaying functions of distance away from the patch location, and show that the amplitude of the perturbations scales linearly with surface slope and patch length while the decay length scales with thickness and patch length, and is critically dependent on the basal boundary condition outside of the slippery patch. These fundamental relationships can be incorporated into more complex models to investigate whether rapid lake drainages on ice sheets, which create a partially lubricated bed, can generate sufficient stress and velocity perturbations in the overlying ice flow to trigger lake drainage cascades.

\end{abstract}

\begin{keywords}
Authors should not enter keywords on the manuscript, as these must be chosen by the author during the online submission process and will then be added during the typesetting process (see \href{https://www.cambridge.org/core/journals/journal-of-fluid-mechanics/information/list-of-keywords}{Keyword PDF} for the full list).  Other classifications will be added at the same time.
\end{keywords}

{\bf MSC Codes }  {\it(Optional)} Please enter your MSC Codes here

%%%%%%%%%%%%%%%%%%%%%%%%%%%%%%%%%%%%%%%%%%%%%%%%%%%
%%%%%%%%%%%%% SECTION 1: INTRODUCTION %%%%%%%%%%%%%
%%%%%%%%%%%%%%%%%%%%%%%%%%%%%%%%%%%%%%%%%%%%%%%%%%%
\section{Introduction}\label{sec:introduction}
%%%%%%%%%%%%%%%%%%%%%%%%%%%%%%%%%%%%%%%%%%%%%%%%%%%
Pressure-driven viscous flow subjected to spatially variable slip boundary conditions are found in settings spanning from micro-scale industry applications \citep[e.g.,][]{lauga_effective_2003,wexler_shear-driven_2015,liu_effect_2016} to geophysical-scale settings \citep[e.g.,][]{vedeneeva_spreading_2021,barcilon_steady_1993,schoof_ice_2007, mantelli_ice_2019,shah_two-layer_2021}.  In the flow of glacial ice in particular, the interfaces between varying boundary conditions are critical to ice sheet evolution under the warming climate.  For example, the organization of ice flow into ice streams, which serve as fast flowing conduits through which Antarctic ice is rapidly transported seaward \citep[e.g.,][]{bindschadler_changes_1998,hulbe_century-scale_2007,rignot_ice_2011}, is dependent on the dynamics associated with a transition in basal boundary condition \citep{jacobson_thermal_1998, suckale_deformation-induced_2014, mantelli_ice_2019}.  The flux of ice through the grounding line (e.g., off the edge of the continent to form floating ice shelves in the ocean) is critically dependent on the dynamics at the basal boundary transition of the grounding line \citep{weertman_stability_1974,chugunov_modelling_1996,schoof_ice_2007}.  Another such glacial interface is a seasonally occurring feature of the Greenland Ice Sheet (GrIS): rapid spatially finite injections of surface meltwater to the ice-bed interface from the drainage of supraglacial lakes through the ice column \citep[e.g.,][]{tsai_model_2010,stevens_greenland_2015}.

During the summer months (June-September) the GrIS ice-bed interface becomes partially lubricated by surface meltwater which penetrates the ice column via surface-to-bed pathways such as moulins and hydrofracture-driven lake drainages (HLD) \citep[e.g.,][]{alley_access_2005,das_fracture_2008}.  These spatially-finite injections of meltwater reduce the friction between the ice and bed below, leading to a transient acceleration of the overlying ice \citep{joughin_influence_2013, stevens_greenland_2016}.  In many cases, the resulting induced stresses are sufficient to trigger further HLD events, delivering further lubricating water to the bed \citep{das_fracture_2008,stevens_greenland_2015,christoffersen_cascading_2018,poinar_challenges_2021}.  These rapid HLD events can deliver large volumes of water ($\sim5\textrm{ x }10^7\textrm{ m}^3$) to the bed \citep{fitzpatrick_decade_2014}, which form spatially-finite blisters of diameter on the order of several ice thicknesses beneath the ice, reducing the effective pressure (ice overburden minus basal water pressure) over that area \citep[e.g.,][]{lai_hydraulic_2021,stevens_elastic_2024}.

In some cases, multiple neighboring lakes are observed to drain in rapid succession leading to the hypothesis of a domino effect of HLD events, where one lake drainage may trigger another and so on, which could have the capacity to then transiently overwhelm the evacuation capacity of the subglacial hydrological system, leading to increased ice motion \citep[e.g.,][]{christoffersen_cascading_2018}.  Prescience of these dynamics is especially relevant as the presence of lakes is projected to expand inland over the coming decades in response to the warming climate \citep{leeson_supraglacial_2015,igneczi_northeast_2016,macferrin_rapid_2019}.  These inland regions are characterized by a subglacial hydrological system which may be more readily overwhelmed by meltwater injections due to thicker more gently sloping ice, potentially leading to a lower effective basal friction in response to rapid basal meltwater injection (e.g., HLD) \citep{chandler_evolution_2013,meierbachtol_basal_2013,dow_upper_2014,doyle_persistent_2014}.  Constraining the structure of the influence one lake drainage has on other lakes in the area via distal stress communication, and how this communication varies across the GrIS, remains a critical yet incomplete challenge.  In this study, we aim to investigate the following question: what are the fundamental controls on the magnitude and spatial extent of stress and velocity response to spatially finite basal traction loss?

Our case sits within a broader class of ice flow problems subjected to spatially varying basal boundary conditions.  For grounded ice sheets this type of transition often occurs at the onset (margins) of narrow rapidly moving regions, known as ice streams, which are fed laterally by slower moving ice ridges that are basally frozen.  \citet{jacobson_thermal_1998} demonstrated the sensitivity of ice stream margin locations to thermo-viscous feedbacks.  \citet{haseloff_boundary_2015} derive a boundary layer description of this lateral transition from slow shear-type to fast plug-type flow, accounting for the role of lateral advection on margin migration.  \citet{barcilon_steady_1993} describe the structure of flow across an along-flow basal condition transition from no-slip to free-slip, work which was extended by \citet{mantelli_ice_2019} and \citet{schoof_role_2021} to include the thermomechanical behavior in the associated boundary layers.  The present paper differes in that we discuss a spatially finite region of rapid flow with an upstream transition from slow to fast flow and a downstream transition back to slow flow.  We are also interested in systems for which the slippery region is not induced or controlled thermomechanically, rather simply by the presence of meltwater which can supply sufficient upward pressure on the overlying ice to drastically reduce basal friction, such as is the case following a rapid lake drainage \citep[e.g.,][]{lai_hydraulic_2021}.

Previous attention has been paid to the problem of ice flow over spatially finite basal heterogeneities.  In particular, several studies demonstrate steady-state longitudinal transmission of basal heterogeneities (e.g., topography) to be on the order of several ice thicknesses \citep[e.g.,][]{kamb_stress-gradient_1986,gudmundsson_transmission_2003,sergienko_glaciological_2013,crozier_basal_2018}.  The lengthscale over which information transmits from the bed to the surface is consistent with the findings in our depth-averaged framework presented in this paper.  We contribute to the understanding of flow over spatially finite heterogeneities by examining explicitly the relationship between observable geometrical parameters (e.g., thickness, slope, patch length) and the resulting flow perturbations.  We reveal, using simple a analytical model and scaling arguments, the dominant controlling relationships among geometrical parameters and the instantaneous perturbed viscous extensional stress and velocity.  In this paper, we do not evolve the system forward in time, as the slippery patch-induced instantaneous stress perturbations are directly relevant for triggering surface fracture and potential subsequent lake drainage cascades. However, long-term impacts of the slippery patch's annual appearance on ice-sheet dynamics is an interesting topic for future exploration.

We begin, in \S\ref{sec:analytical_model}, by presenting an analytical model based on the shallow stream approximation (SSA). We then present a comparison between this simplified analytical model and the results of numerical simulations for a range of rheologies in \S\ref{sec:numerical_simulations}.  We conclude our work in \S\ref{sec:glaciological_implications} with a discussion on implications of the work for the GrIS under the warming climate.

%%%%%%%%%%%%%%%%%%%%%%%%%%%%%%%%%%%%%%%%%%%%%%%%%%%
%%%%%%%%%%%%% SECTION 2: 1D DA MODEL %%%%%%%%%%%%%
%%%%%%%%%%%%%%%%%%%%%%%%%%%%%%%%%%%%%%%%%%%%%%%%%%%
\section{Analytical model}\label{sec:analytical_model}
\begin{figure}
    \centering
    \includegraphics[width=1.0\linewidth]{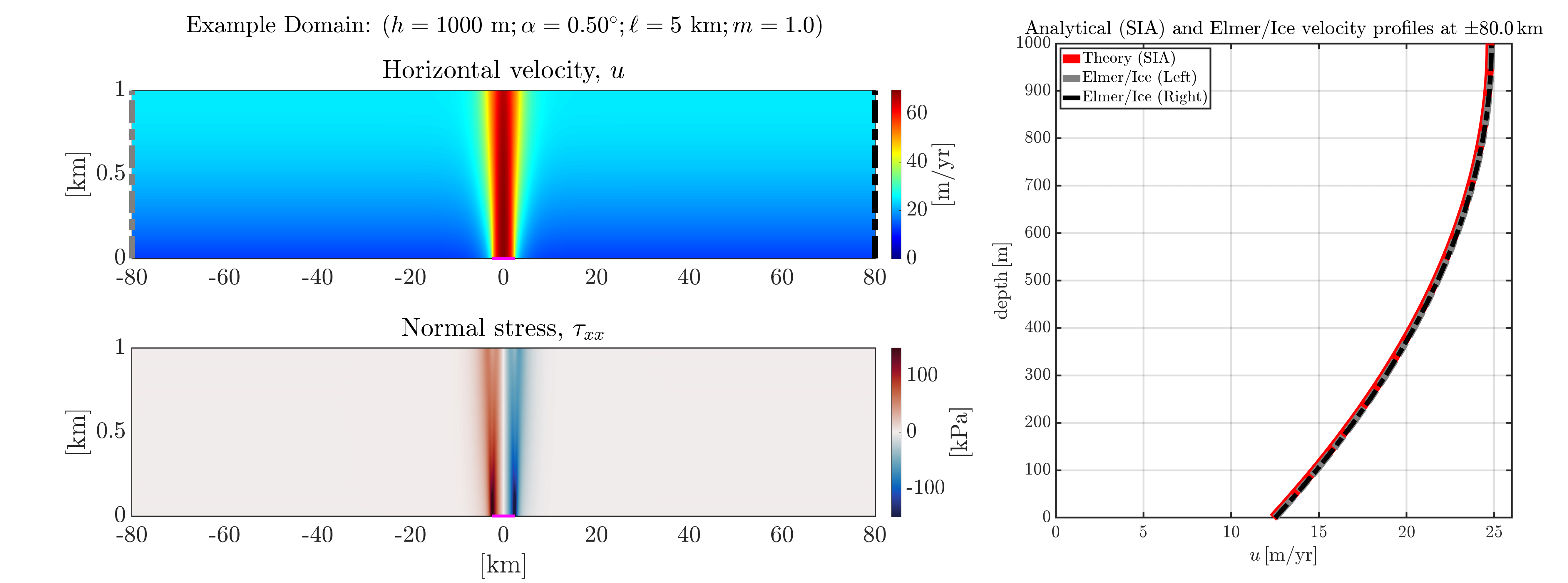}
    \caption{Example numerical domain displaying the horizontal velocity (top) and normal deviatoric stress (bottom) result of an numerical experiment run with Elmer/Ice.  Boundary conditions are set sufficiently far from the patch location such that the velocity in the far field is consistent with the sliding-accommodated shallow ice approximation (right) and $\tau_{xx}$ returns to zero.}
    \label{fig:domain_example}
\end{figure}

%%%%%%%%%%%%%%%%%%%%%%%%%%%%%%%%%%%%%%%%%%%%%%%%%%%
We here develop a model considering an along-flow two-dimensional ice sheet in a vertical plane $(x,z)$ of constant thickness, $h$, and slope $\alpha$, where $x$ is along the horizontal and $z$ the vertical direction (figure \ref{fig:domain_example}).  The ice-bed interface is characterized by high shear stress, $\tau_b$, everywhere except for a free-slip region of finite length, $\ell$, henceforward called the `slippery patch' or simply the `patch' (pink region in figure \ref{fig:domain_example}).  We consider the well-known shallow stream approximation (SSA) \citep[e.g.,][]{macayeal_large-scale_1989} describing depth-averaged ice flow:
\begin{align}\label{eqn:ssa}
    2(h\tau_{xx})_x - \tau_b = \rho g h \partial_xh \, ,
\end{align}
where the depth-averaged horizontal extensional deviatoric stress, $\tau_{xx}$, is tied to the depth-averaged horizontal velocity $u$ and its along-flow horizontal gradient $u_x$ via the rheology $\tau_{xx} = 2\eta u_x$. We use a power law viscosity given by $\eta = \frac{1}{2}A^{-1/n}\dot{\epsilon_e}^{(1-n)/n}$, where $\dot{\epsilon}_e$ is the second invariant of the strain rate tensor and we assume $\dot{\epsilon}_e\approx u_x$.  Basal shear stress is set here to follow a sliding law of the general form
\begin{equation}\label{eqn:taub_DA}
\tau_b = Cu^m \, .
\end{equation}
We assume constant thickness, $h$, and constant slope of $\partial_xh = -\alpha$, reducing our governing equation \eqref{eqn:ssa} to:
\begin{align}\label{eqn:governing_equation}
    2hA^{-1/n} (u_{xx})^{1/n} - Cu^m = -\rho g h \alpha \, .
\end{align}
We are particularly interested in a solution in terms of elementary functions, which will simply reveal the structure of stress and velocity perturbations. As such, we first solve for the simplified case of linear sliding and Newtonian flow (e.g., $m = 1/n=1$). The value of the sliding coefficient, $C$, is determined by the relative contributions of basal sliding and internal deformation to the overall ice velocity.  In particular, we define a dimensionless parameter, $\gamma$, to capture this ratio of the scale of the shearing-velocity contributions, $[u_{shear}]$, to the scale of sliding-velocity contributions, $[u_{slide}]$:
\begin{equation}\label{eqn:gamma}
    \gamma \equiv \frac{[u_{shear}]}{[u_{slide}]} = \frac{A\tau_b^nh}{(\tau_b/C)^{(1/m)}} = Ah\tau_b^{(n-\frac{1}{m})}C^{\frac{1}{m}}\, .
\end{equation}
Rearranging equation \ref{eqn:gamma}, making the substitution for $\tau_b = \rho g \alpha h$, gives:
\begin{equation}\label{eqn:C_def}
    C \equiv \gamma^m (\rho g \alpha)^{(1-mn)} A^{-m} h^{(1-m-mn)}\, .
\end{equation}
Substituting equation \ref{eqn:C_def} into equation \ref{eqn:governing_equation}, we have
\begin{equation}
    (u_{xx})^\frac{1}{n} - \frac{\gamma^m (\rho g \alpha)^{(1-mn)}A^{(\frac{1}{n}-m)}h^{-m(1+n)}}{2}u^m = -\frac{A^\frac{1}{n}\rho g \alpha}{2} \, ,
\end{equation}
which simplifies significantly upon the assumption of $m = \frac{1}{n} = 1$:
\begin{equation}\label{eqn:governing_equation_analytical}
    u_{xx} - \frac{\gamma}{2h^2}u = -\frac{A\rho g \alpha}{2} \, .
\end{equation}
Equation \ref{eqn:governing_equation_analytical} can be solved analytically to give the stress and velocity responding to the presence of the slippery patch, including the stress and velocity fields upstream from, across the length of, and downstream from the patch.  To start, we examine the perturbation structure across the length of the patch (\textsection\ref{sec:anmodel_patch}), as we will use this as boundary conditions to later determine the upstream (downstream) stress and velocity perturbation structures (\textsection\ref{sec:anmodel_outside}).  Later on in \textsection\ref{sec:numerical_simulations} we use numerical experiments to extend the work to more general cases (e.g., $m\neq\frac{1}{n}\neq1$) and analyze the results through the lens provided by the analytical structure presented here.

%%%%%%%%%%%%%%%%%%%%%%%%%%%%%%%%%%%%%%%%%%%%%%%%%%%
\subsection{Stress and velocity across the slippery patch for all $n, m$}\label{sec:anmodel_patch}
The region inside the slippery patch is subjected to no basal shear stress (i.e., $\gamma=0$), reducing our depth-averaged governing equation (equation \ref{eqn:ssa}) to
\begin{align}
    (\tau_{xx})_x = -\frac{\rho g \alpha}{2} \, ,
\end{align}
which, integrating and applying an anti-symmetry condition of vanishing stress at the patch center, yields
\begin{align}\label{eqn:soln_txx_patch}
    \tau_{xx} = -\frac{\rho g \alpha}{2}x\, , \text{ for } \left[ -\frac{\ell}{2} \leq x \leq \frac{\ell}{2} \right] \, .
\end{align}
At the upstream end of the patch $x=-\ell/2$, the value of stress is given as the pressure drop across the patch due to the surface slope and patch length, scaled by a factor of $1/4$:
\begin{equation}\label{eqn:BC_txx_transpt}
    \tau_{xx}(x=-\ell/2) = \frac{\rho g \ell \alpha}{4} \, ,
\end{equation}
revealing the natural scale for the perturbed stress magnitude.  Rescaling the equation by the relevant scales for stress magnitude, $[\tau_{xx}]\equiv \rho g \ell \alpha/4$, and patch length, $\ell$, gives us the non-dimensional form: 
\begin{align}\label{eqn:stress_patch_nondim}
    \tilde{\tau}_{xx} = -2\tilde{x} &\text{ where } \tilde{x}\equiv\frac{x}{\ell}, \tilde{\tau}_{xx}\equiv\frac{\tau_{xx}}{\rho g\ell\alpha/4}  \text{ for } [-1/2 \leq \tilde{x} \leq 1/2] \,.
\end{align}
Integrating \ref{eqn:stress_patch_nondim} along $\tilde{x}$ with a specified rheology will yield the depth-averaged velocity across the slippery patch. For Newtonian fluids with a linear sliding law ($m=\frac{1}{n}=1$), integration of \ref{eqn:stress_patch_nondim} along $\tilde{x}$ reveals a quadratic velocity structure within the patch, 
\begin{equation}\label{eqn:u_patch_observation}
    \tilde{u} = -\tilde{x}^2 + c_1 \, ,
\end{equation}
where determination $c_1$, and therefore the specific form of the velocity profile, depends on the basal conditions (sliding law) outside of the patch (see \textsection\ref{sec:anmodel_outside_velo}).

%%%%%%%%%%%%%%%%%%%%%%%%%%%%%%%%%%%%%%%%%%%%%%%%%%%
\subsection{Stress and velocity outside of the slippery patch; cases with $n=\frac{1}{m}=1$}\label{sec:anmodel_outside}
The stress and velocity perturbation structures in the upstream (downstream) regions outside the slippery patch are determined by using the depth-averaged stress profile across the patch as a boundary condition (equation \ref{eqn:BC_txx_transpt}).  To do so analytically, we consider cases of Newtonian ice, $\tau_{xx}=A^{-1} u_x$, and linear Weertman-type sliding law, $\tau_b = Cu^m$ with $m=\frac{1}{n}=1$, beginning with a solution for stress (\textsection\ref{sec:anmodel_outside_stress}) and then velocity (\textsection\ref{sec:anmodel_outside_velo}).  We demonstrate the solutions for the upstream regions but report also the downstream solutions, which take the same form save for an inverted sign.
%%%%%%%%%%%%%%%%%%%%%%%%%%%%%%%%%%%%%%%%%%%%%%%%%%%
\subsubsection{Stress outside the patch ($n=\frac{1}{m}=1$)}\label{sec:anmodel_outside_stress}
Solving equation \ref{eqn:governing_equation_analytical}, subjected to stress continuity at the patch onset at $x=-\ell/2$ (equation \ref{eqn:BC_txx_transpt}), yields an analytical solution for the upstream stress perturbation:
\begin{align}\label{eqn:soln_txx_bl_sliding}
    \tau_{xx} = \frac{\rho g \alpha \ell}{4}\exp{\left( \frac{\sqrt{\gamma}}{h\sqrt{2}}(x+\ell/2)\right)}\, , \text{ for } \left[ -\infty\leq x \leq -\frac{\ell}{2} \right] \, .
\end{align}
The scale for the stress, $[\tau_{xx}] \equiv \rho g \alpha \ell / 4$, is the same for the upstream region as for across the patch (equation \ref{eqn:stress_patch_nondim}). Outside the patch, however, the natural scale for the horizontal coordinate, $x$, is now $h/\sqrt{\gamma}$.  Thus, across the full domain, the dimensionless depth-averaged stress is summarized as:
\begin{equation}\label{eqn:summary_nondim_txx_sliding}
\tilde{\tau}_{xx}(\tilde{x}) = 
    \begin{cases}
        \exp{\left( \frac{1}{\sqrt{2}} \left(\tilde{x}^* + \frac{\lambda\sqrt{\gamma}}{2}\right) \right)} &\text{ where } \tilde{x}^*=\frac{x}{(h/\sqrt{\gamma})}  \text{ for } [- \infty \leq \tilde{x}^* \leq -\lambda\sqrt{\gamma}/2]\, \\
        -2\tilde{x} &\text{ where } \tilde{x}=\frac{x}{\ell}  \text{ for } [-1/2 \leq \tilde{x} \leq 1/2]\, \\
        -\exp{\left( -\frac{1}{\sqrt{2}} \left(\tilde{x}^* - \frac{\lambda\sqrt{\gamma}}{2}\right) \right)} &\text{ where } \tilde{x}^*=\frac{x}{(h/\sqrt{\gamma})}  \text{ for } [\lambda\sqrt{\gamma}/2 \leq \tilde{x}^* \leq \infty]\, ,
    \end{cases}
\end{equation}
where $\lambda\equiv\ell/h$ is the ratio of patch length to thickness, and $\gamma$ is the ratio of shear to sliding velocity contributions (equation \ref{eqn:gamma}).  Here, we use $\tilde{x}^*$ to denote the dimensionless horizontal coordinate outside the patch.

%%%%%%%%%%%%%%%%%%%%%%%%%%%%%%%%%%%%%%%%%%%%%%%%%%%
\subsubsection{Velocity outside the patch ($n=\frac{1}{m}=1$)}\label{sec:anmodel_outside_velo}
In the upstream region, integration of equation \ref{eqn:soln_txx_bl_sliding} gives the depth-averaged horizontal velocity
\begin{align}
    u(x) = \frac{A\rho g \alpha\ell h}{4}\frac{\sqrt{2}}{\sqrt{\gamma}}\exp{\left( \frac{\sqrt{\gamma}}{h\sqrt{2}}(x+\ell/2) \right)} + u_L \, , \text{ for } \left[ -\infty\leq x \leq -\frac{\ell}{2} \right] \, ,
\end{align}
where $u_L$ is the integration constant fixed by satisfying the depth averaged SIA velocity at $x\to-\infty$ (see Appendix \ref{sec:appendixA_bcs_and_uL}).  We focus just on the velocity perturbed above background, $u_p(x) \equiv u(x) - u_L$, which is
\begin{align}\label{eqn:up_sliding}
    u_p(x) = \frac{A\rho g \alpha\ell h}{4}\frac{\sqrt{2}}{\sqrt{\gamma}}\exp{\left( \frac{\sqrt{\gamma}}{h\sqrt{2}}(x+\ell/2) \right)} \, , \text{ for } \left[ -\infty\leq x \leq -\frac{\ell}{2} \right] \, .
\end{align}
Across the patch, integration of equation \ref{eqn:soln_txx_patch} and enforcement of velocity continuity at $x=-\ell/2$ gives:
% \begin{align}
%     u_p(x) = -A\frac{\rho g \alpha}{4}x^2 + c_2 \, , \text{ for } \left[ -\frac{\ell}{2} \leq x \leq \frac{\ell}{2} \right] \,,
% \end{align}
% where velocity continuity requires:
% \begin{align}
%     &\frac{A\rho g \alpha\ell}{4}\sqrt{2}h = -A\frac{\rho g \alpha}{16}\ell^2 + c_2\\
%     \Rightarrow& c_2 = \frac{A\rho g \alpha \ell h}{4}\left( \sqrt{2} + \frac{\ell}{4h} \right) \, ,
% \end{align}
% which gives
\begin{align}\label{eqn:up_dimfull}
    u_p(x) = \frac{A\rho g \alpha\ell h}{4}\frac{1}{\sqrt{\gamma}}\left( -\frac{x^2\sqrt{\gamma}}{\ell h} + \sqrt{2} + \frac{\ell\sqrt{\gamma}}{4h} \right) \text{ for } \left[ -\frac{\ell}{2} \leq x \leq \frac{\ell}{2} \right] \,.
\end{align}
Rescaling gives the dimensionless velocity across the domain:
\begin{equation}\label{eqn:summary_nondim_u_sliding}
\hspace{-1cm}\tilde{u}_p(\tilde{x}) = 
    \begin{cases}
        \sqrt{2}\exp{\left( \frac{1}{\sqrt{2}}\left( \tilde{x}^* + \frac{\lambda\sqrt{\gamma}}{2} \right) \right)} &\text{ where } \tilde{x}^*\equiv\frac{x}{(h/\sqrt{\gamma})}, \tilde{u}_p\equiv\frac{u_p}{(A\rho g \alpha\ell h/4\sqrt{\gamma})} \text{ for } [-\infty \leq \tilde{x}^{*} \leq -\lambda\sqrt{\gamma}/2]\, \\ 
        -\tilde{x}^{{**}^2} &\text{ where } \tilde{x}^{**}\equiv\frac{x}{\sqrt{\ell h}}, \tilde{u}_p^{**}\equiv\frac{u_p}{A\rho g \alpha\ell h/4} - \frac{\lambda}{4} - \frac{\sqrt{2}}{\sqrt{\gamma}}  \text{ for } \left[-\sqrt{\frac{\lambda}{4}} \leq \tilde{x}^{**} \leq \sqrt{\frac{\lambda}{4}} \right] \\
        \sqrt{2}\exp{\left( -\frac{1}{\sqrt{2}}\left( \tilde{x}^* - \frac{\lambda\sqrt{\gamma}}{2} \right) \right)} &\text{ where } \tilde{x}^*\equiv\frac{x}{(h/\sqrt{\gamma})}, \tilde{u}_p\equiv\frac{u_p}{(A\rho g \alpha\ell h/4\sqrt{\gamma})} \text{ for } [\lambda\sqrt{\gamma}/2 \leq \tilde{x}^{*} \leq \infty]\, ,\\
    \end{cases}
\end{equation}
consistent with the quadratic form in equation \ref{eqn:u_patch_observation}. Comparing equation \ref{eqn:summary_nondim_u_sliding} with the stress perturbation length scales (equation \ref{eqn:summary_nondim_txx_sliding}), the velocity outside the slippery patch follows the same decay length scale as that of the stress, $h/\sqrt{\gamma}$, and hence is rescaled to the same dimensionless length $\tilde{x}^*$. Within the slippery patch, however, the velocity varies horizontally with a length scale $\sqrt{\ell h}$, differing from the stress variation length scale, $\ell$, which is thickness-independent (equation \ref{eqn:summary_nondim_txx_sliding}). Hence we use $\tilde{x}^{**}$ to denote the new dimensionless velocity variation length scale within the slippery patch. The velocity scale inside the patch differs from that outside the patch, hence our use of $\tilde{u}_p^{**}$ to denote the dimensionless velocity inside the patch.

\subsection{Sliding ratio $\gamma$}
The value of $\gamma$ depends on the ratio of shear to sliding velocity contributions outside the slippery patch (equation \ref{eqn:gamma}; see also \textsection\ref{sec:sims_an_num_comp}). In cases which allow for some sliding in the regions outside of the patch, albeit with significant basal friction, the result may be demonstrated using a more precise boundary layer analysis approach (Appendix \ref{sec:appendixC_asymptotics}). In the next section, we use numerical simulations to explore the stress and velocity perturbations associated with a range of parameter values for $h, \alpha, \ell, n, m, \textrm{ and } \gamma$, further generalizing our results to nonlinear rheology ($n\neq1$) and sliding behavior ($m\neq1$).  For the comparisons between the numerical results and our analytical model, we first discuss the relationship between the two types of basal stress parameterizations used.  Namely, in the analytical model, we have used $\tau_b = Cu^m$ (equation \ref{eqn:taub_DA}), which gives the shear stress in terms of the depth-averaged velocity, $u$.  In contrast, the numerical experiments use
\begin{equation}\label{eqn:taub_b}
    \tau_b = C_bu_b^m \, ,
\end{equation}
where $C_b$ is the basal sliding coefficient we feed to the numerical simulations. Equation \ref{eqn:taub_b} relates to the parameterization used in the analytical model (equation \ref{eqn:taub_DA}) via:
\begin{equation}\label{eqn:C_vs_Cb}
    C = C_b\left( \frac{u_b}{u} \right)^m \, .
\end{equation}
The ratio of sliding velocity, $u_b$ to total depth-averaged velocity, $u$, can be written, (e.g., see equation \ref{eqn:SIA}), as
\begin{equation}\label{eqn:ub_u}
    \frac{u_b}{u} = \frac{(\rho g h \alpha)^{1/m} C_b^{-1/m}}{\frac{2A}{n+2}(\rho g h \alpha)^nh + (\rho g h \alpha)^{1/m}C_b^{-1/m}} \, .
\end{equation}
In order to draw comparisons between the scaling of the coupling length and the numerical experiments (e.g., \textsection\ref{sec:numerical_simulations}), we must first return to the definition of $\gamma$ (equation \ref{eqn:gamma}), noticing that it is defined in terms of $C$, but $C_b$ is what we have fed to the Elmer simulations.  To bridge this gap, we substitute equation \ref{eqn:C_vs_Cb} into equation \ref{eqn:gamma} to obtain
\begin{equation}\label{eqn:gamma_def2}
    \gamma = Ah(\rho g h \alpha)^{n-1/m}C_b^{1/m}\left( \frac{u_b}{u} \right)\, ,
\end{equation}
where the ratio $\left( \frac{u_b}{u} \right)$ is given in equation \ref{eqn:ub_u}. 

\subsubsection{Cases of $n=\frac{1}{m}=1$}
For comparisons between the analytical model and the cases where $m=\frac{1}{n}=1$, equation \ref{eqn:ub_u} simplifies to
\begin{equation}
    \frac{u_b}{u} = \frac{3}{2AhC_b + 3} \, ,
\end{equation}
and equation \ref{eqn:gamma_def2} becomes
\begin{equation}
    \gamma_{an} = \frac{3AhC_b}{2AhC_b + 3} \, ,
\end{equation}
where the $[*]_{an}$ subscript denotes that this applies to the analytical model. To parameterize the end-member no slip (outside of the slippery patch) cases into our SSA-based analytical model, we effectively increase the value of $C_b \to \infty$, which eliminates basal velocity, i.e., $u_b\to0$.  In this case, we observe the following limit
\begin{equation}
    \gamma_{ns} = \lim_{C_b\to\infty}\gamma = \frac{3}{2}\, ,
\end{equation}
where the $[*]_{ns}$ subscript stands for `no slip' (outside of the slippery patch).  This serves as the upper limit value used for the analytical model corresponding to the no slip cases in the numerical simulations.

%%%%%%%%%%%%%%%%%%%%%%%%%%%%%%%%%%%%%%%%%%%%%%%%%%%
%%%%%%%%% SECTION 3: NUMERICAL %%%%%%%%%%%
%%%%%%%%%%%%%%%%%%%%%%%%%%%%%%%%%%%%%%%%%%%%%%%%%%%
\section{Numerical simulations}\label{sec:numerical_simulations}
%%%%%%%%%%%%%%%%%%%%%%%%%%%%%%%%%%%%%%%%%%%%%%%%%%%
We used the finite element full-Stokes solver Elmer/Ice to conduct a suite of 2D along-flow numerical experiments \citep{gagliardini_capabilities_2013}. We depth-averaged the results of these numerical experiments to first validate our analytical model in \textsection\ref{sec:sims_an_num_comp}.  We then extend the analysis to investigate the stress and velocity perturbation structures for cases with nonlinear sliding law (e.g., $m\neq 1$) and rheologies (e.g., Glen's flow law, $n=3$) as depth-averaged quantities and at the ice surface in \textsection\ref{sec:sims_extended}, generalizing the above-derived scaling relationships to these cases.

%%%%%%%%%%%%%%%%%%%%%%%%%%%%%%%%%%%%%%%%%%%%%%%%%%%
\subsection{Comparison between numerical experiments and analytical model ($m=\frac{1}{n}=1$)}\label{sec:sims_an_num_comp}
\begin{figure}
    \centering
    \includegraphics[width=1.0\linewidth]{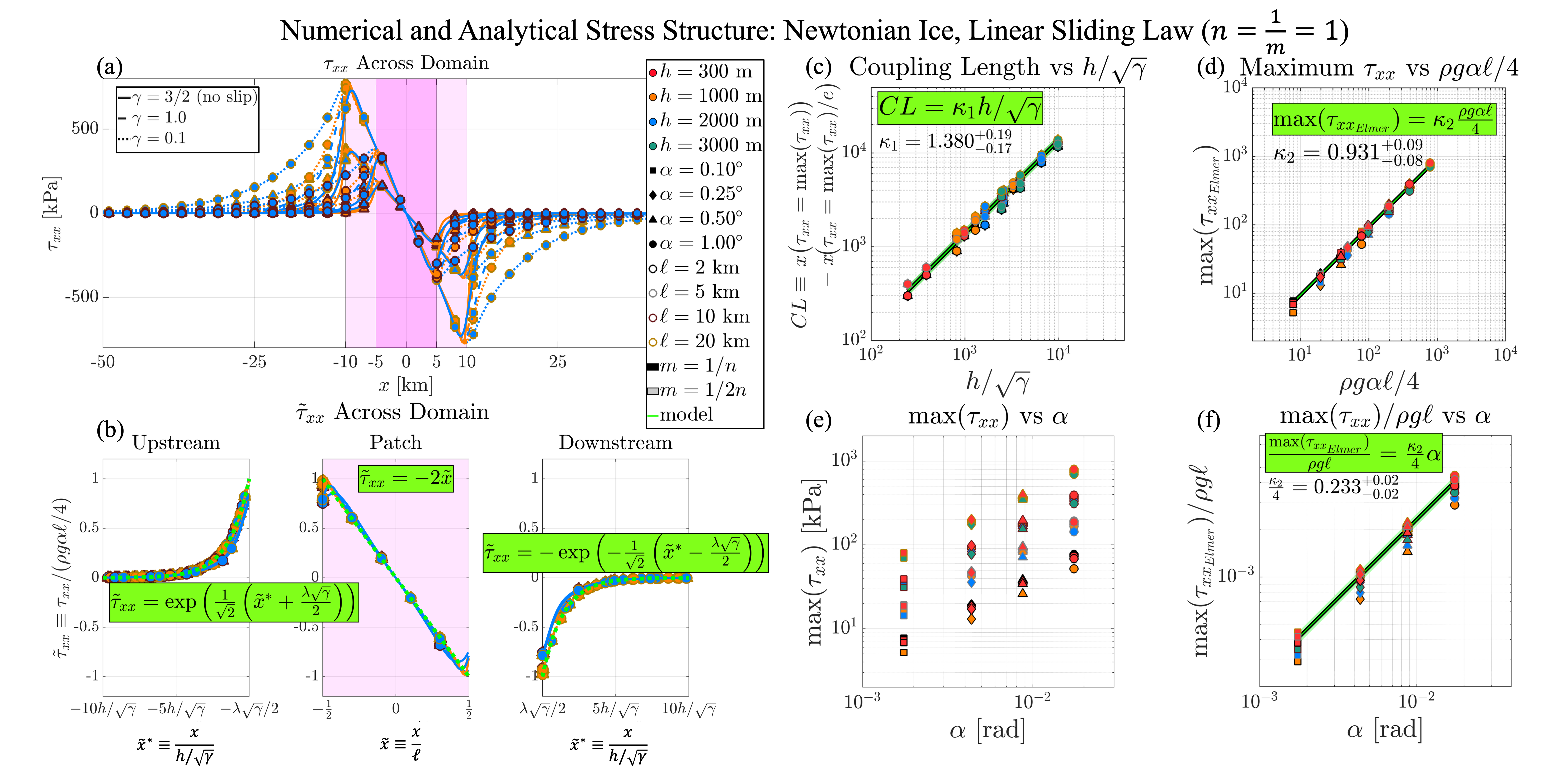}
    \caption{Depth-averaged numerical results for a selection of the numerical experiments with Newtonian ice and linear sliding law outside the patch (e.g., $n=1, m=\frac{1}{n}$) for direct comparison to analytical model, showing (a) the perturbed stress across the domain, (b) non-dimensionalized stresses across the domain in the upstream, patch, and downstream regions, compared to the analytical model (green), and scaling relationships and regression fits (green) for (c) coupling length and (d-f) stress. The solid green lines in (c-f) are the best fits, with the shaded areas representing the $\pm1\sigma$ envelope of the data scatter.}
    \label{fig:newt_stress_annumcomp}
\end{figure}

For this set of experiments, we used the same flow parameters as in the analytical model, namely Newtonian ice ($n=1$, $\eta = 10^{14}$ Pa s) and basal boundary condition outside of the slippery patch characterized by no-slip ($\gamma=3/2$) or by some amount of sliding where $\gamma=[0.1, 1.0]$, with ranges in surface slope, $\alpha = [0.1^\circ, 0.25^\circ, 0.5^\circ, 1.0^\circ]$, thickness, $h = [300\textrm{ m, } 1000\textrm{ m, } 2000\textrm{ m, } 3000\textrm{ m}]$, and patch length, $\ell = [2\textrm{ km, } 5\textrm{ km, } 10\textrm{ km, } 20\textrm{ km}]$.  These simulations yield stress perturbations for a range of $\gamma, \alpha, h, \ell$ that collapse well in the non-dimensional coordinates $\tilde{x}, \tilde{x}^*, \tilde{\tau}_{xx}$ (figure \ref{fig:newt_stress_annumcomp}a,b), showing good agreement with the analytical model (equation \ref{eqn:summary_nondim_txx_sliding}, green curves in figure \ref{fig:newt_stress_annumcomp}b).  All symbols in all panels are denoted by the same legend in panel a. The scaling relationships as revealed by the analytical model (equation \ref{eqn:summary_nondim_txx_sliding}) hold for the numerical results, where the stress magnitude scales linearly with slope and patch length (figure \ref{fig:newt_stress_annumcomp}f), and the spatial extent of the stress perturbation scales (the ``coupling lengthscale" $CL$) linearly with the ice thickness divided by the square root of the sliding ratio, $\gamma$ (equation \ref{eqn:gamma}, figure \ref{fig:newt_stress_annumcomp}c). In order to generate this coupling length ($CL$) comparison between the numerical experiments and the analytical scaling, we define the numerical coupling length ($CL$) to be distance between the location of maximum stress, max$(\tau_{xx})$ (near the transition point at the upstream patch edge), and the upstream location at which the stress decays to below $1/e$ of the maximum value (figure \ref{fig:newt_stress_annumcomp}c).  This reflects the exponential nature of the stress decay in the analytical model. The numerical coupling lengths ($CL$) show good agreement with the theoretical coupling lengths ($h/\sqrt{\gamma}$) (figure \ref{fig:newt_stress_annumcomp}c), and can be expressed as $CL = \kappa_2h/\sqrt{\gamma}$ where $\kappa_2=1.380^{+0.19}_{-0.17}$ is a numerical prefactor fitted across a range of $\gamma, \alpha, h, \ell$ for which $\lambda > 2$. We restrict ourselves to this limit of $\lambda$ as we expect our analysis to break down in the limit as the length of the patch approaches the ice thickness (see Appendix \ref{sec:appendixB_patchlen_choice}).  

The maximum $\tau_{xx}$ in the simulation is slightly less than that in the analytical model, $\rho g \ell \alpha/4$ (equation \ref{eqn:summary_nondim_txx_sliding}, figure \ref{fig:newt_stress_annumcomp}b,d). This is expected, as the analytical model follows from the SSA, which does not capture higher-order dynamics within the boundary layers between the patch and the far-field flow. That said, we observed that the numerical maximum extensional stress (vertical axis in figure \ref{fig:newt_stress_annumcomp}d) follows the same scaling as the analytical model and can be expressed as max$(\tau_{xx})=\kappa_1 \rho g \ell \alpha/4$, where $\kappa_1\approx 0.931^{+0.09}_{-0.09}$ is a numerical prefactor fitted across a range of $\gamma, \alpha, h, \ell$ for which $\lambda > 2$ (figure \ref{fig:newt_stress_annumcomp}e).  Thus, while the analytical depth-averaged stress expression (equation \ref{eqn:summary_nondim_txx_sliding}) slightly overestimates the maximum stress, this overestimation is consistent across all experiments and captured by the constant numerical prefactor, $\kappa_1$.  The analytical model (equation \ref{eqn:summary_nondim_txx_sliding}) thus sufficiently provides the scaling, i.e. the parameter dependence, of the (1) maximum stress ($\approx \rho g \ell \alpha /4$) and the (2) coupling length scale ($\approx h/\sqrt{\gamma}$) with good predictive power across a range of $\gamma, \alpha, h, \ell$.

\begin{figure}
    \centering
    \includegraphics[width=1.0\linewidth]{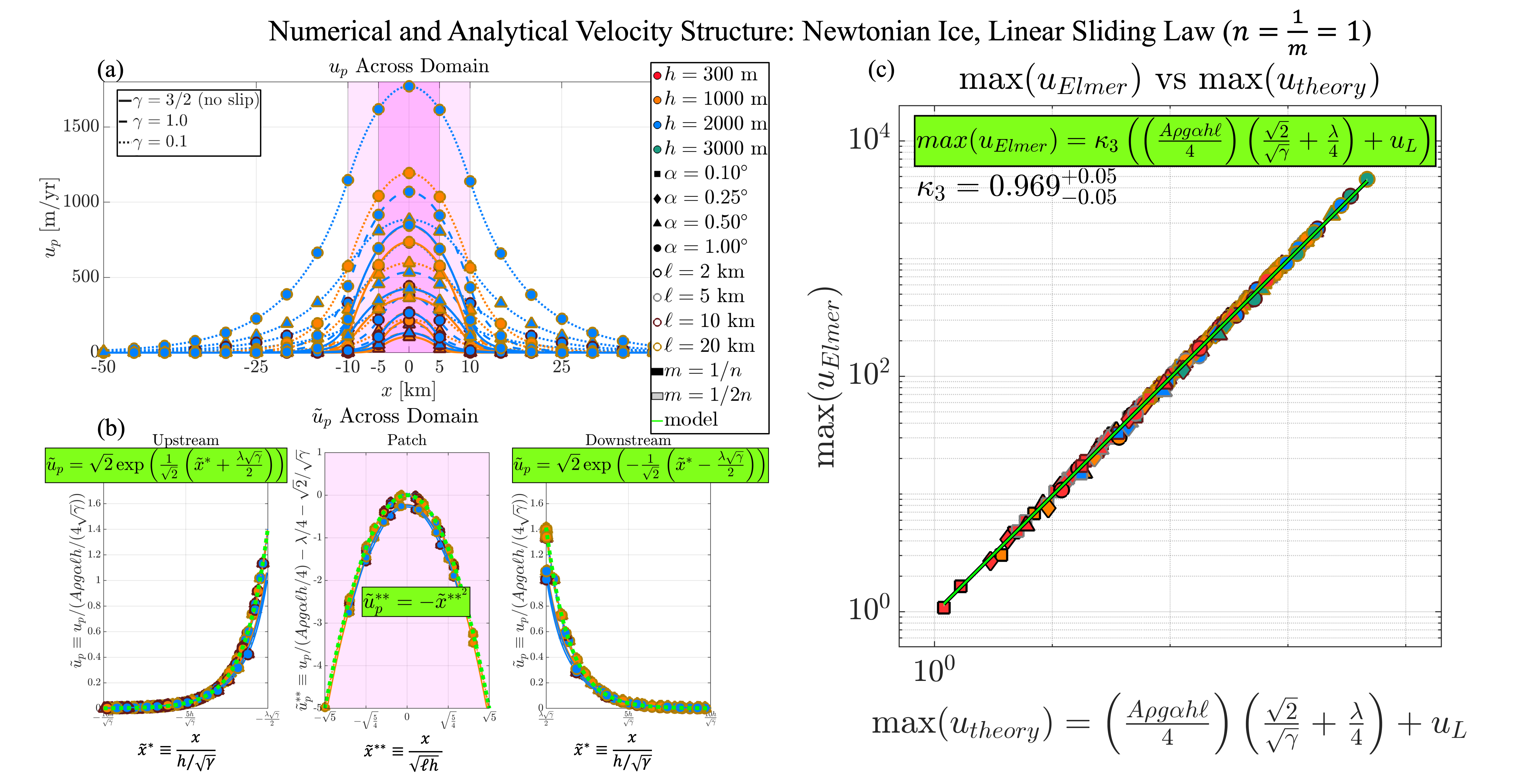}
    \caption{Depth-averaged numerical results for a selection of the numerical experiments with Newtonian ice and linear sliding law outside the patch (e.g., $n=1, m=\frac{1}{n}$) for direct comparison to analytical model, showing (a) the perturbed velocity across the domain, (b) non-dimensionalized velocity across the domain in the upstream, patch, and downstream regions, compared to the analytical model (green), and (c) scaling relationship and regression fit (green) for the maximum velocity magnitude. The solid green line in (c) is the best fit, with the shaded area representing the $\pm1\sigma$ envelope of the data scatter.}
    \label{fig:newt_velo_annumcomp}
\end{figure}

Similarly, the velocity perturbations extracted from these same numerical experiments collapse in the dimensionless coordinates (figure \ref{fig:newt_velo_annumcomp}a,b) and agree well with the analytical results (equation \ref{eqn:summary_nondim_u_sliding}, green curves in figure \ref{fig:newt_velo_annumcomp}b) across the range of $\gamma, \alpha, \ell, h$. The maximum perturbed velocity exhibits excellent agreement with the analytical prediction (figure \ref{fig:newt_velo_annumcomp}c), providing a scaling relationship showing the parameter dependence of the maximum velocity as max$(u_{elmer})=\kappa_3(A\rho g \alpha \ell h/4)(\sqrt{2/\gamma} + \lambda/4) + u_L$, with a numerical prefactor $\kappa_3=0.969^{+0.05}_{-0.05}$.  The decay of stress outside the patch is exponential (equation \ref{eqn:summary_nondim_txx_sliding}), and therefore so is the velocity decay (equation \ref{eqn:summary_nondim_u_sliding}), giving decay lengthscale $(\approx h/\sqrt{\gamma})$, as expected.

The analytical model best describes cases that are subjected to some amount of slip outside the patch region (e.g., $\gamma<3/2$; no-slip corresponds to $\gamma=3/2$), which is to be expected due to the model's basis in SSA.  To fully capture the dynamics in the transition zone adjacent to the slippery patch (especially for the no-slip to free-slip cases, solid curves, figures \ref{fig:newt_stress_annumcomp}, \ref{fig:newt_velo_annumcomp}) requires a full-Stokes analysis in two dimensions \citep[e.g.,][]{barcilon_steady_1993, chugunov_modelling_1996, mantelli_ice_2019}.  In utilizing the SSA to model the problem at hand, our aim is to identify the simplest model that reveals the dominant structure of stress and velocity perturbations in response to a slippery patch. The full boundary layer analysis is left for future work.

%%%%%%%%%%%%%%%%%%%%%%%%%%%%%%%%%%%%%%%%%%%%%%%%%%%
\subsection{Comparison between numerical experiments and scaling arguments (general $n, m$)}\label{sec:sims_extended}
Here, we extend the simulations to include a broader range of cases.  Namely, we use a sliding law characterized by an exponent deviating from the Weertman-type exponent, relaxing the requirement of $m=\frac{1}{n}$, and present the stress and velocity structure across the surface, in addition to depth-averaged.  The scaling relationships from the theory are extended as well.  Namely, the scale for the stress magnitude remains the same $(\approx \rho g \alpha \ell/4)$, but the scale for the coupling length (equation \ref{eqn:xbl_ext}) and maximum velocity (equation \ref{eqn:veloscale_ext}) are now dependent upon the sliding law exponent, $m$, and the stress exponent, $n$.  We reach this conclusion through the following scaling argument. Let's denote the region of interest, which sits upstream from the onset of the slippery patch, as the boundary layer, where the velocity and horizontal extent occupied by this boundary layer are denoted as $[u_{bl}]$ and $[x_{bl}]$, respectively.  The upstream ice in this boundary layer, which is subjected to extensional stress, $[\tau_{xx}]=\rho g \alpha\ell/4$, set by the presence of the patch, is characterized by the sliding law with $C \equiv \gamma^m (\rho g \alpha)^{(1-mn)} A^{-m} h^{(1-m-mn)}$ (equation \ref{eqn:C_def}).  Combining the assumption of a power-law rheology, $[u_{bl}]/[x_{bl}] = A[\tau_{xx}]^n$, with a balance between extensional stress and basal shear in the transition region between upstream ice and the patch, $h[\tau_{xx}]/[x_{bl}] = C[u_{bl}]^m$, gives
\begin{equation}
    h/[\tau_{xx}][x_{bl}] = C(A[\tau_{xx}^n][x_{bl}])^m\, .
\end{equation}
Substituting for $[\tau_{xx}]$ and $C$ and rearranging gives us our boundary layer (thus the coupling length) scale:

\begin{equation}\label{eqn:xbl_ext}
    [x_{bl}] \equiv \gamma^{-\frac{m}{m+1}} h^{\frac{m+mn}{m+1}} \left( \frac{\ell}{4} \right)^{\frac{1-mn}{m+1}} \, ,
\end{equation}
which reduces to $[x_{bl}]=h/\sqrt{\gamma}$ when $m=\frac{1}{n}=1$, consistent with our analytical model (e.g., 
\textsection\ref{sec:analytical_model}, equation \ref{eqn:summary_nondim_txx_sliding}). We now generalize the maximum velocity scale following the same procedure used to determine equation \ref{eqn:up_dimfull}.  In particular, integration of the stress structure across the domain while using equation \ref{eqn:xbl_ext} in place of the analytical scale $(h/\sqrt{\gamma})$ gives
\begin{equation}\label{eqn:veloscale_ext}
    [\textrm{max}(u_{theory})] = A\left(\frac{\rho g \alpha\ell}{4}\right)^n \left( \frac{\gamma^{-\frac{m}{m+1}} h^{\frac{m+mn}{m+1}} \left( \frac{\ell}{4} \right)^{\frac{1-mn}{m+1}}}{n} + \frac{\ell}{2(n+1)} \right) + u_L \, .
\end{equation}
We use these generalized scales for coupling length (equation \ref{eqn:xbl_ext}) and maximum perturbed velocity (equation \ref{eqn:veloscale_ext}) to extend our analysis first to Newtonian ice with nonlinear sliding laws and then to non-Newtonian ice.

\subsubsection{Newtonian ice with nonlinear sliding $n = 1, m= \frac{1}{n}, \frac{1}{2n}$}
\begin{figure}
    \centering
    \includegraphics[width=1.0\linewidth]{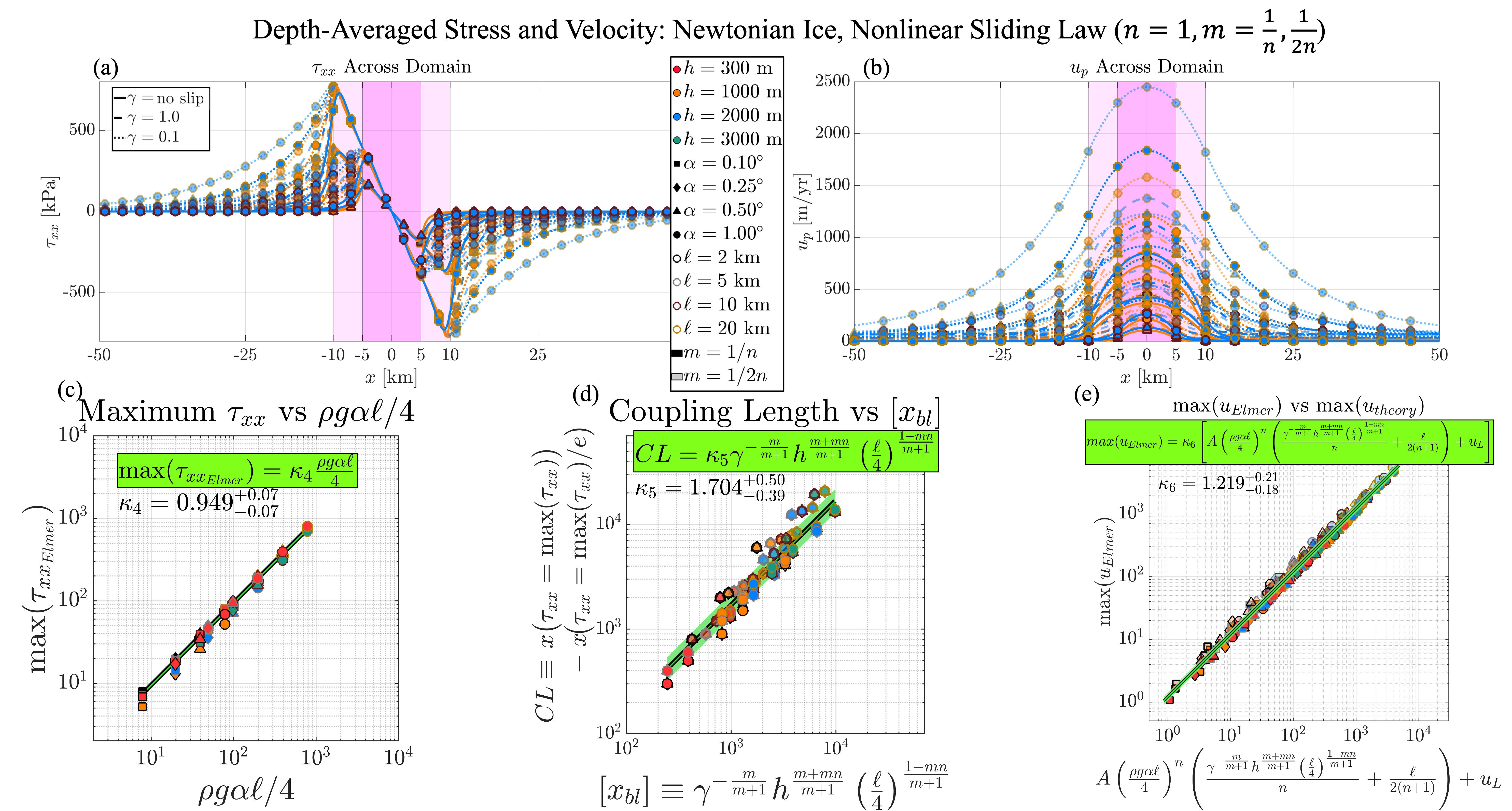}
    \caption{Depth-averaged numerical results for a selection of the numerical experiments with Newtonian ice extended to include nonlinear sliding law outside the patch (e.g., $n=1, m=\frac{1}{n},\frac{1}{2n}$) showing the perturbed stress (a) and velocity (b) across the domain.  Scaling relationships for stress magnitude (c), coupling length (d), and velocity magnitude (e) are consistent with the scaling arguments in equations \ref{eqn:xbl_ext} and \ref{eqn:veloscale_ext}. The solid green lines are the best fits, with the shaded areas representing the $\pm1\sigma$ envelope of the data scatter.}
    \label{fig:newt_extDA}
\end{figure}

To generalize our analysis, we begin by extending to the cases which consider a nonlinear sliding law exponent but still Newtonian ice, namely $m=\frac{1}{2n}=\frac{1}{2}$.  This serves two purposes: the first is demonstrating the utility of our scaling analyses beyond cases with direct analogy to the analytical model. The second is an illustration that in addition to the sliding ratio (e.g., $\gamma$), the form of the sliding law (e.g., $m$-value) influences the stress and velocity perturbation structure.  As a first step in this demonstration, we begin by depth-averaging numerical experiments across the same range of $\gamma, \alpha, h, \ell$ as in \textsection\ref{sec:sims_an_num_comp}, but including cases with $m=\frac{1}{2}$ (figure \ref{fig:newt_extDA}).  We find the following. (1) The dominant scaling relationships hold, consistent with our framework presented in \textsection\ref{sec:analytical_model}. The stress magnitude scales with patch length and slope, max$(\tau_{xx}) = \kappa_4\rho g\ell\alpha/4$, where the numerical prefactor now $\kappa_4=0.949^{+0.07}_{-0.07}$ (figure \ref{fig:newt_extDA}c). (2) The coupling lengthscale ($CL$) is again a function of sliding ratio and thickness, but $m\neq1/n$ introduces the influence of the patch length $\ell$, meaning that for arbitrary $n, m$, the coupling lengthscale is a weighted geometric mean of sliding ratio, thickness, and patch length (equation \ref{eqn:xbl_ext}, figure \ref{fig:newt_extDA}d).  In particular, we see $CL=\kappa_5\gamma^{-\frac{m}{m+1}} h^{\frac{m+mn}{m+1}} \left( \frac{\ell}{4} \right)^{\frac{1-mn}{m+1}}$, with $\kappa_5=1.704^{+0.50}_{-0.39}$.  (3) Because coupling lengthscale is now a factor in the maximum velocity, the theoretical maximum velocity scale is also now modified by this additional factor of patch length (figure \ref{fig:newt_extDA}e).  Namely, $\textrm{max}(u_{elmer})=\kappa_6[A\left(\frac{\rho g \alpha\ell}{4}\right)^n \left( \frac{\gamma^{-\frac{m}{m+1}} h^{\frac{m+mn}{m+1}} \left( \frac{\ell}{4} \right)^{\frac{1-mn}{m+1}}}{n} + \frac{\ell}{2(n+1)} \right) + u_L]$, with $\kappa_6=1.219^{+0.21}_{-0.18}$.

\begin{figure}
    \centering
    \includegraphics[width=1.0\linewidth]{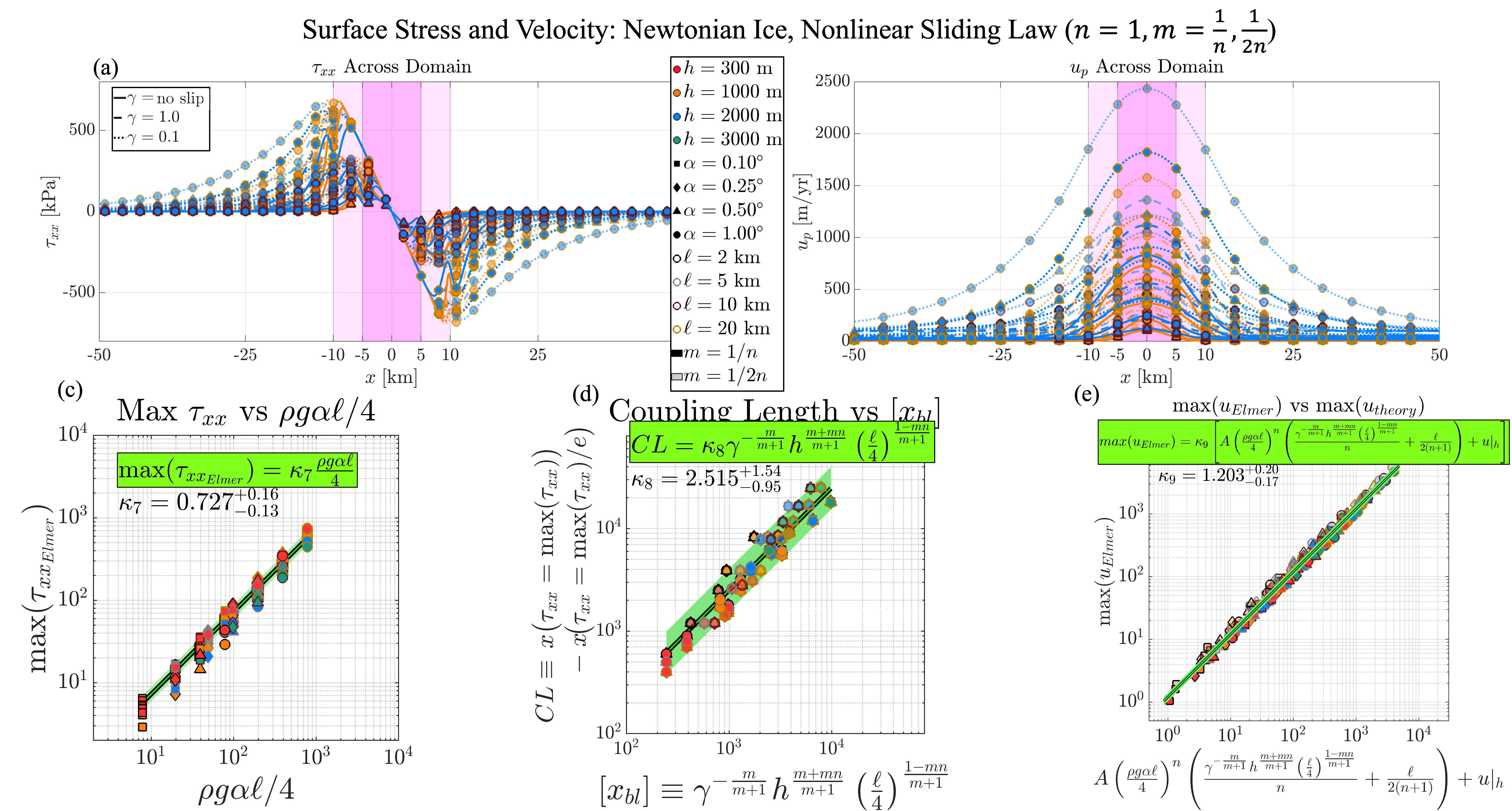}
    \caption{Ice-surface numerical results for a selection of the numerical experiments with Newtonian ice extended to include nonlinear sliding law outside the patch (e.g., $n=1, m=\frac{1}{n},\frac{1}{2n}$) showing the perturbed stress (a) and velocity (b) across the domain.  Scaling relationships for stress magnitude (c), coupling length (d), and velocity magnitude (e) are consistent with the scaling arguments in equations \ref{eqn:xbl_ext} and \ref{eqn:veloscale_ext}. The solid green lines are the best fits, with the shaded areas representing the $\pm1\sigma$ envelope of the data scatter.}
    \label{fig:newt_extSURF}
\end{figure}

Though our analytical model, and the resulting scaling relationships presented above, were developed for depth-averaged flow, we are also interested in an investigation of the stress and velocity structure at the ice surface, as those quantities can be easily observed in the field or lab experiments.  We do so now, noting that the same scaling relationships may be extended, albeit with differences in numerically-determined prefactors.  In particular, the relationships for stress magnitude, coupling length, and velocity maximum are as follows (figure \ref{fig:newt_extSURF}): 
\begin{align}
\textrm{max}(\tau_{{xx}_{surf}}) =& \kappa_7\rho g\alpha\ell/4, \kappa_7=0.727^{+0.16}_{-0.13}\\ 
CL_{surf} =& \kappa_8\gamma^{-\frac{m}{m+1}} h^{\frac{m+mn}{m+1}} \left( \frac{\ell}{4} \right)^{\frac{1-mn}{m+1}}, \kappa_8=2.515^{+1.54}_{-0.95}\\
\textrm{max}(u_{surf}) =& \kappa_{9}\left[A\left(\frac{\rho g \alpha\ell}{4}\right)^n \left( \frac{\gamma^{-\frac{m}{m+1}} h^{\frac{m+mn}{m+1}} \left( \frac{\ell}{4} \right)^{\frac{1-mn}{m+1}}}{n} + \frac{\ell}{2(n+1)} \right) + u_L\right], \kappa_{9}=1.023^{+0.20}_{-0.17} \, .
\end{align}

\subsubsection{Non-Newtonian ice with nonlinear sliding $n = 3, m= \frac{1}{n},\frac{1}{2n}$}
\begin{figure}
    \centering
    \includegraphics[width=1.0\linewidth]{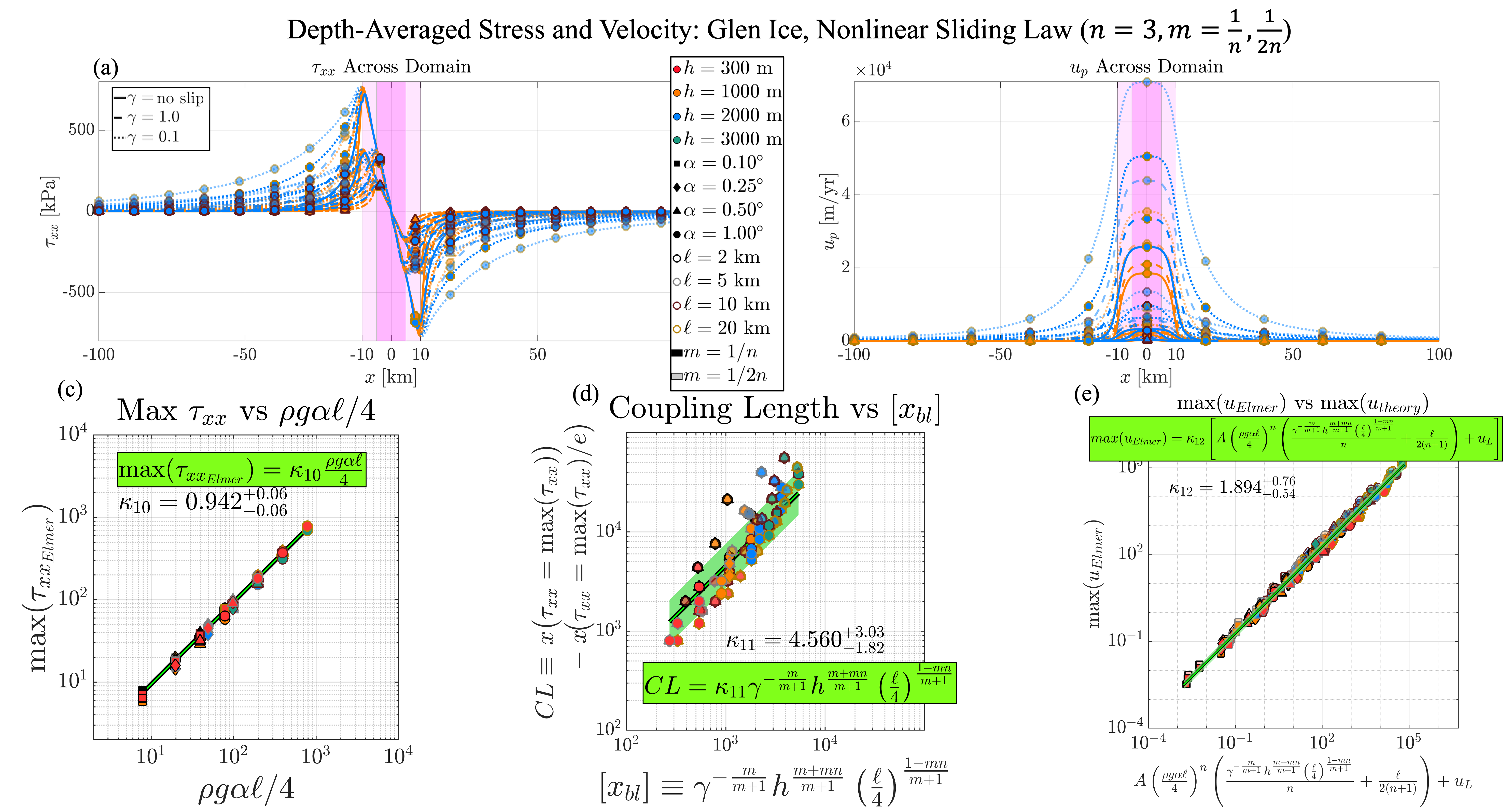}
    \caption{Depth-averaged numerical results for a selection of the numerical experiments with Glen ice including nonlinear sliding law outside the patch (e.g., $n=3, m=\frac{1}{n},\frac{1}{2n}$) showing the perturbed stress (a) and velocity (b) across the domain.  Scaling relationships for stress magnitude (c), coupling length (d), and velocity magnitude (e) are consistent with the scaling arguments in equations \ref{eqn:xbl_ext} and \ref{eqn:veloscale_ext}. The solid green lines are the best fits, with the shaded areas representing the $\pm1\sigma$ envelope of the data scatter.}
    \label{fig:glen_extDA}
\end{figure}

\begin{figure}
    \centering
    \includegraphics[width=1.0\linewidth]{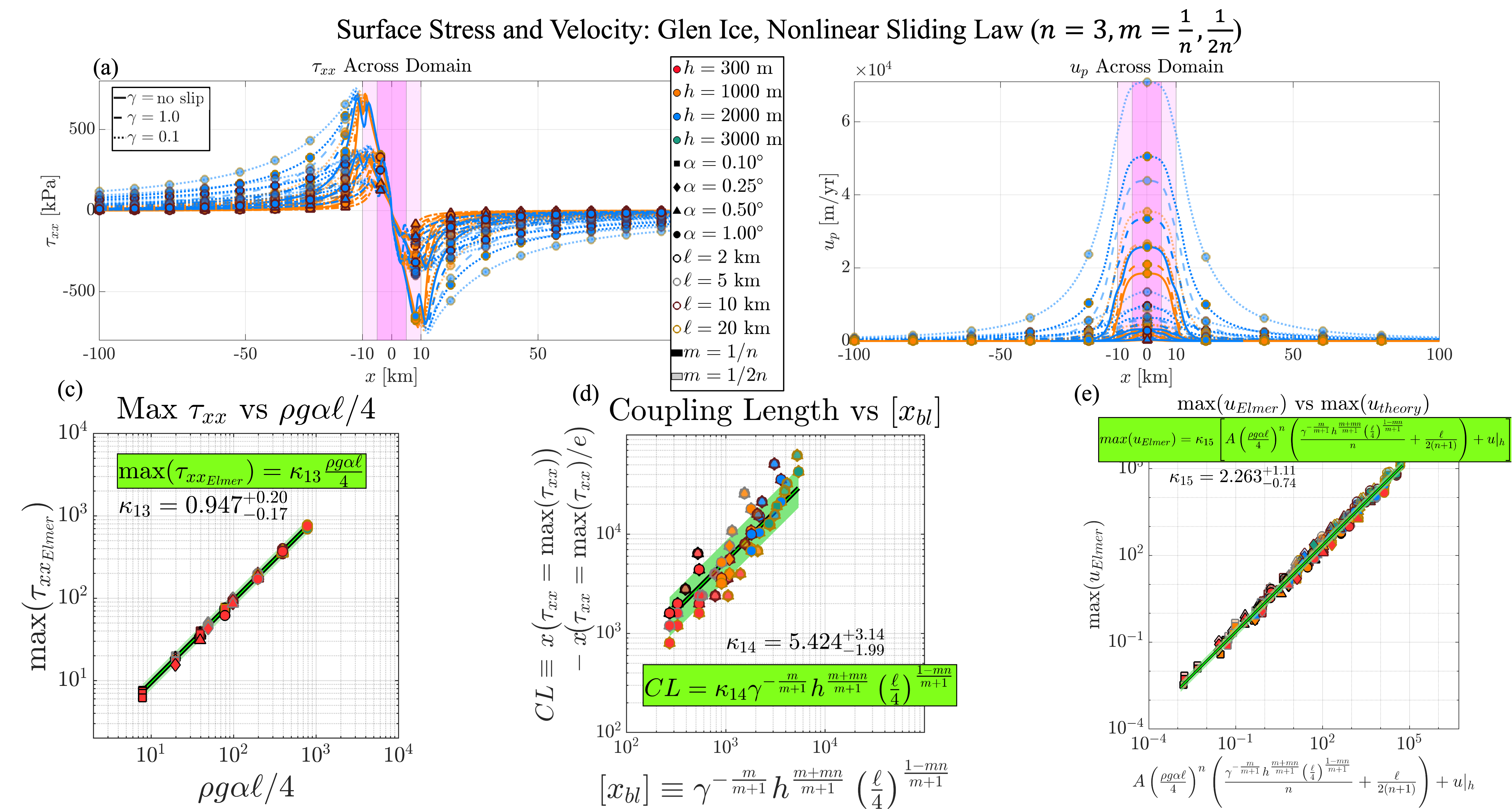}
    \caption{Ice-surface numerical results for a selection of the numerical experiments with Glen ice including nonlinear sliding law outside the patch (e.g., $n=3, m=\frac{1}{n},\frac{1}{2n}$) showing the perturbed stress (a) and velocity (b) across the domain. Scaling relationships for stress magnitude (c), coupling length (d), and velocity magnitude (e) are consistent with the scaling arguments in equations \ref{eqn:xbl_ext} and \ref{eqn:veloscale_ext}. The solid green lines are the best fits, with the shaded areas representing the $\pm1\sigma$ envelope of the data scatter.}
    \label{fig:glen_extSURF}
\end{figure}

In addition to the numerical experiments run with linear ice rheology, we ran a suite of experiments across the same parameter grid but for non-Newtonian ice.  Namely, we used a Glen law rheology with $n=3$, $m=\frac{1}{n}, \frac{1}{2n}$, and $A=3.5\textrm{ x }10^{-25}\textrm{ Pa}^{-3}\textrm{s}^{-1}$ \citep{cuffey_paterson_2010}.  The scaling relationships discussed above extend to these more general cases, as summarized in figures \ref{fig:glen_extDA} \ref{fig:glen_extSURF}.  In particular, the stress magnitude is set by the length of the free-slip patch and the surface slope (figures \ref{fig:glen_extDA}c, \ref{fig:glen_extSURF}c), the coupling length is controlled by a weighted geometric mean among the sliding ratio, thickness, and patch length (figures \ref{fig:glen_extDA}d, \ref{fig:glen_extSURF}d), and the maximum velocity across the patch is controlled by a combination of the coupling length, far-field velocity, patch length, and surface slope (figures \ref{fig:glen_extDA}e, \ref{fig:glen_extSURF}e).  These relationships hold for the depth-averaged results (figure \ref{fig:glen_extDA}) as well as at the ice surface (figure \ref{fig:glen_extSURF}), albeit with different numerical prefactors.  For these experimental results, we restrict ourselves to cases with a patch length to thickness ratio of at least $5$ (see Appendix \ref{sec:appendixB_patchlen_choice}). A complete list of scaling arguments and the numerical prefactors for all relevant numerical experiments are summarized in table \ref{tab:num_prefactors}. The fact that all numerical prefactors are close to order of magnitude one shows that the scaling relationships in table \ref{tab:num_prefactors} can provide good order-of-magnitude estimates of the maximum perturbed stress, velocity, and coupling length scale, in addition to capturing their dependence on a range of $\gamma, \alpha, h, \ell $.

\begin{table}
  \centering
  \caption{Numerically determined prefactors, $\kappa$, for scaling relationships. DA and S denote depth-averaged and surface, respectively}
  \label{tab:num_prefactors}
  \hspace*{-1cm}
  \begin{tabular}{|c|c|c|c|}
    \toprule
    \makecell[c]{Case}
      & \makecell[c]{$\tau_{{xx}_{max}}=$ \\ $\kappa[\rho g\alpha\ell/4]$}
      & \makecell[c]{$CL=$ \\ \scriptsize{$\kappa\left[\gamma^{-\tfrac{m}{m+1}}\,h^{\tfrac{m+mn}{m+1}}\,( \tfrac{\ell}{4} )^{\tfrac{1-mn}{m+1}}\right]$}}
      & \makecell[c]{$u_{max}=$ \\ \scriptsize{$\kappa\bigl[A\left(\frac{\rho g \alpha\ell}{4}\right)^n \left( \frac{\gamma^{-\frac{m}{m+1}} h^{\frac{m+mn}{m+1}} \left( \frac{\ell}{4} \right)^{\frac{1-mn}{m+1}}}{n} + \frac{\ell}{2(n+1)} \right) + u_L]$}} \\
    \toprule
    \toprule
    $n,m=1$ (DA)  & $\kappa_2 = 0.931^{+0.09}_{-0.09}$ & $\kappa_1 = 1.380^{+0.19}_{-0.17}$ & $\kappa_3 = 0.969^{+0.05}_{-0.05}$ \\ 
    \midrule
    $n=1,m=\tfrac{1}{n},\tfrac{1}{2n}$ (DA) & $\kappa_4 = 0.949^{+0.07}_{-0.07}$ & $\kappa_5 = 1.704^{+0.50}_{-0.39}$ & $\kappa_6 = 1.219^{+0.21}_{-0.18}$\\ 
    \midrule
    $n=1,m=\tfrac{1}{n},\tfrac{1}{2n}$ (S) & $\kappa_7 = 0.727^{+0.16}_{-0.13}$ & $\kappa_8 = 2.515^{+1.54}_{-0.95}$ & $\kappa_{9} = 1.203^{+0.20}_{-0.17}$\\ 
    \midrule
    $n=3,m=\tfrac{1}{n},\tfrac{1}{2n}$ (DA) & $\kappa_{10} = 0.942^{+0.06}_{-0.06}$ & $\kappa_{11} = 4.560^{+3.03}_{-1.82}$ & $\kappa_{12} = 1.894^{+0.76}_{-0.54}$\\ 
    \midrule
    $n=3,m=\tfrac{1}{n},\tfrac{1}{2n}$ (S) & $\kappa_{13} = 0.947^{+0.20}_{-0.17}$ & $\kappa_{14} = 5.424^{+3.14}_{-1.99}$ & $\kappa_{15} = 2.263^{+1.11}_{-0.74}$\\ 
    \bottomrule
  \end{tabular}
\end{table}

%%%%%%%%%%%%%%%%%%%%%%%%%%%%%%%%%%%%%%%%%%%%%%%%%%%
%%%%% SECTION 4: GLACIOLOGICAL IMPLICATIONS %%%%%%
%%%%%%%%%%%%%%%%%%%%%%%%%%%%%%%%%%%%%%%%%%%%%%%%%%%
\section{Implications}\label{sec:glaciological_implications}
The analysis in this paper can be broadly applied to scenarios in nature and engineering involving gravity-driven viscous flow over a partially lubricated solid surface. Here, we discuss its specific implications for ice sheets.
An open question facing the glaciology community is: how are stresses coupled between neighboring regions across the Greenland Ice Sheet (GrIS)? This question stems from the need to understand how the ice sheet generally will respond to continued arctic warming and meltwater forcing, with implications ranging from the dynamics of summer HLD speed up events to hydrofracture cascades \citep[e.g.,][]{christoffersen_cascading_2018}.  For example, \citet{zwally_surface_2002} observed an increase in summer ice surface velocities at a location $\sim 40$ km inland from the western GrIS margin (Swiss Camp), attributing this speed-up to the presence of meltwater at the ice-bed interface due to HLD events. It was later shown, however, that the speed-up observed at this location could be described by an alternative mechanism.  Namely, \citet{price_seasonal_2008} demonstrated that a seasonal acceleration near the coast ($>$10 km downstream from Swiss Camp) may have caused the observed speed-up.

The communication of stresses between lakes in response to their drainage is still yet to be fully constrained.  It remains unclear precisely how and how far these stresses may be transmitted in response to HLD-induced spatially finite meltwater injections to the bed.  Modeling of membrane stress response to basal traction loss by \citet{christoffersen_cascading_2018} suggests there may be a mechanism by which lakes located several tens of kilometers apart may transmit stresses and provide a triggering mechanism among themselves.  Observations of elastic stresses generated by HLD events by \citet{stevens_elastic_2024} suggest a coupling length on the order of several ice thicknesses (a few kilometers).  In this study, we contribute to this conversation by concluding that our simple model helps reveal the fundamental structure of stress perturbations in response to the basal slip induced by a lake drainage.  In particular, we conclude that the magnitude of stress perturbations is set by the pressure drop across the slippery patch, $\rho g\alpha\ell$ (\textsection\ref{sec:anmodel_patch}).  The perturbed stress extends (decays) into the upstream ice over a lengthscale set by a weighted geometric mean of the sliding ratio, thickness, and patch length, and is critically dependent on the form of sliding outside the patch (\textsection\ref{sec:anmodel_outside_stress}).

\begin{figure}
    \centering
    \includegraphics[width=1.0\linewidth]{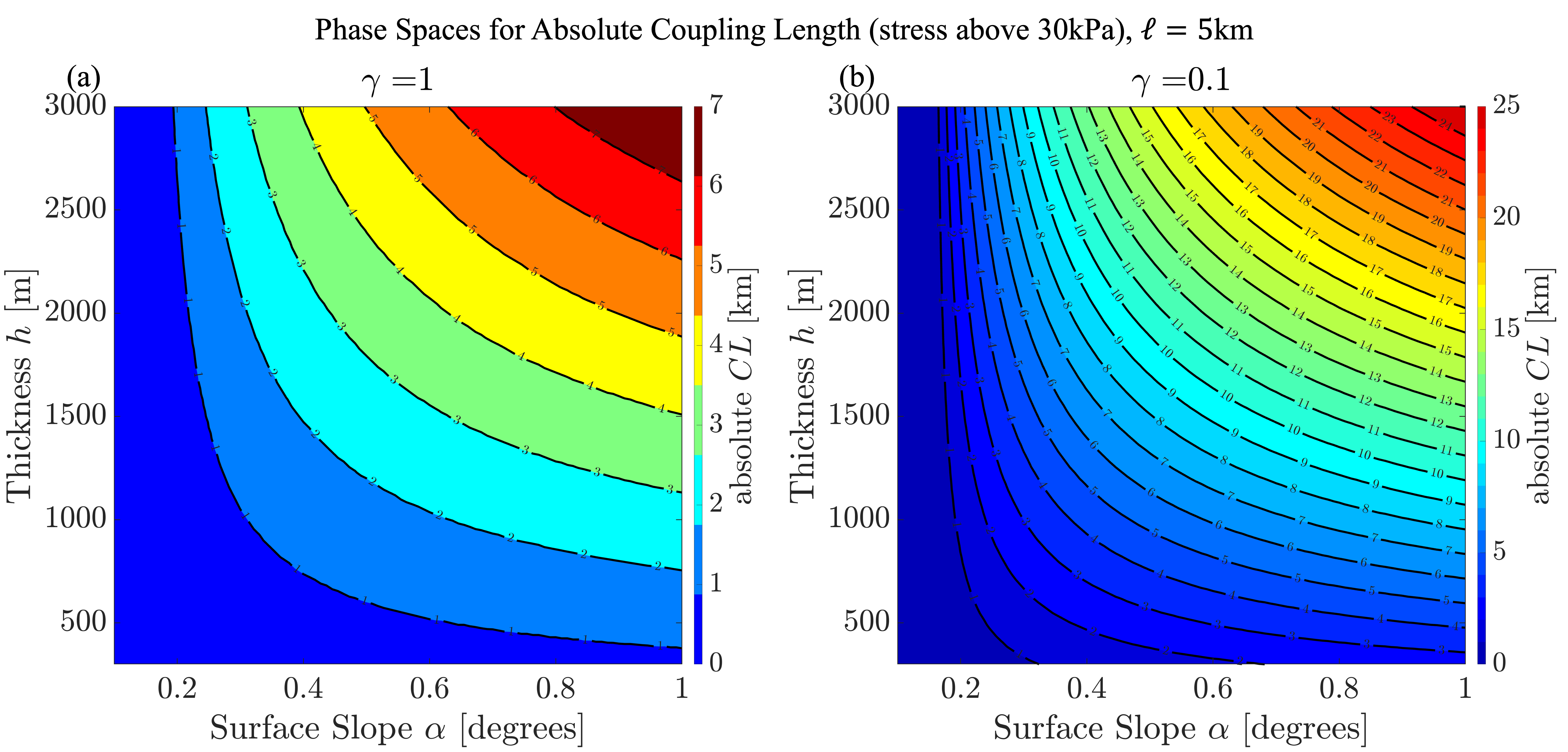}
    \caption{Phasespace for coupling length above threshold of 30 kPa as a function of thickness $h$ and surface slope $\alpha$ for a bed characterized by a sliding ratio of $\gamma=1$ (a) and $\gamma=0.1$ (b).  The term `absolute coupling length' here is meant to indicate the actual distance over which stresses are perturbed above a certain threshold value (here chosen to be 30 kPa), as opposed to the scale of the stress decay, which we previously refer to as coupling length $(CL)$.}
    \label{fig:phasespace_newt_analytic}
\end{figure}

\begin{figure}
    \centering
    \includegraphics[width=1.0\linewidth]{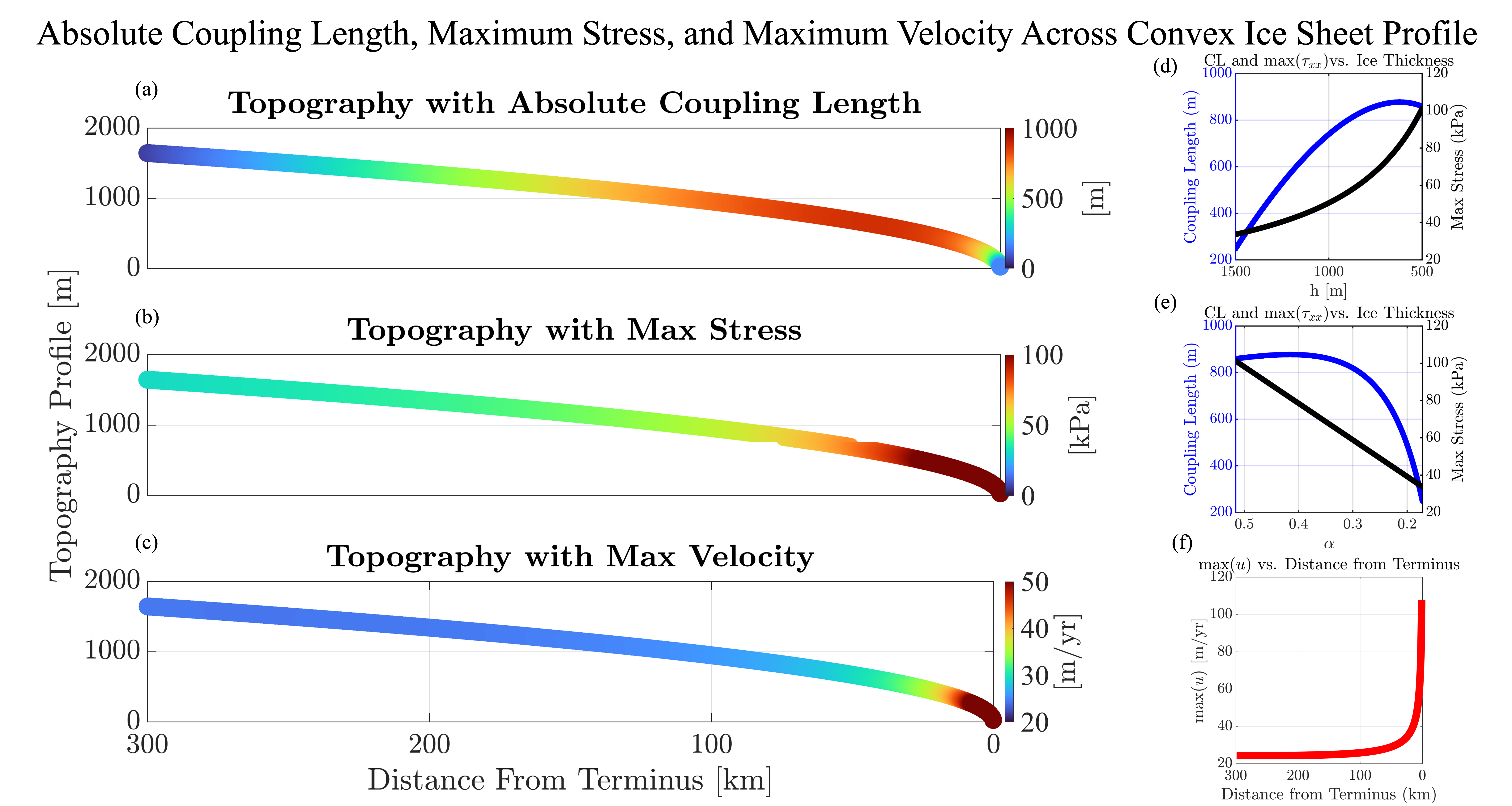}
    \caption{Results from our analytical model (assuming $\gamma=1$, $\ell=5$ km) for coupling length (a), stress magnitude (b), and velocity magnitude (c) painted on a simplified parabolic ice sheet topography profile \citep[e.g.,][]{weertman_milankovitch_1976}. Coupling length and stress magnitude dependence on thickness and surface slope are shown in (d) and (e), respectively. Velocity perturbation decays with distance from terminus (f). Thus, inland regions exhibit a weaker response to the presence of a slippery patch compared to regions closer to the terminus due to the flatter surface slope and therefore weaker pressure drop across the patch.  This principle holds generally for any convex ice topography profile.}
    \label{fig:topopaint}
\end{figure}

Though the stress coupling length increases with the local ice thickness, the exponential nature of the stress response suggests that we should be particular in our definition of the term `coupling length'.  Say, for example, we are interested in the absolute distance over which the perturbed upstream stress is above a given value.  As an illustrative example, we here use 30 kPa due to the onset of fracture initiation \citep[e.g.,][]{lai_vulnerability_2020}.  This extent, referred to in the discussion below as the `absolute coupling length', is then a function not only of the stress-decay lengthscale but also the maximum magnitude of the stress perturbation.  The maximum stress magnitude, and therefore the `absolute coupling length', are then unique functions of location on the ice sheet.  From our analytical model, we computed the maximum stress magnitude and absolute coupling length (upstream distance over which the perturbed stress is above 30 kPa) at each combination of surface slope and thickness, assuming a slippery patch length of 5 km with two choices of sliding ratio, $\gamma=0.1, 1$ (figure \ref{fig:phasespace_newt_analytic}). Projecting these values onto the ice sheet profile reveals that, for a convex surface topography \citep[e.g.,][]{halfar_dynamics_1981, weertman_milankovitch_1976}, locations further upstream (far from the terminus) experience smaller stress perturbations and shorter absolute coupling lengths (figure \ref{fig:topopaint}).

The influence of the sliding behavior outside of the patch region is made clear in the comparison between two phasespaces that show the absolute coupling length for a range of thickness and surface slopes while fixing patch length at 5 km (figure \ref{fig:phasespace_newt_analytic}).  Cases with a larger sliding ratio, $\gamma$, (figure \ref{fig:phasespace_newt_analytic}a) incur reduced absolute coupling lengths compared with cases with smaller $\gamma$ (figure \ref{fig:phasespace_newt_analytic}b) across the range of thicknesses and slopes (note the difference in colorbar between figure \ref{fig:phasespace_newt_analytic}a,b).  For example, at a surface slope of $0.5^\circ$ and a thickness of 1000 m, the absolute coupling length varies from $\sim$ 1.5 km for $\gamma=1$ to $\sim$ 4.5 km for $\gamma=0.1$.  This implies that not only is the amount of spatially finite meltwater injection (e.g., patch length) important in controlling stress perturbations, but so too is the amount of `background' slipperiness into which this meltwater injection is placed.

Thus, though the presence of supraglacial lakes is expected to continue expanding inland from the GrIS margin under the warming future climate \citep[e.g.,][]{leeson_supraglacial_2015}, these inland locations may be more shielded from the same hydrofracture cascade effects currently modeled and observed for lakes closer to the margin.  Stress perturbations may be of smaller magnitude, leading to shorter absolute coupling lengths despite thicker ice, due to flatter surface slopes.  Lakes then may be less likely to drain via hydrofracture, and the ones that do may be less likely to trigger cascades with nearby lakes.  Further constraints on these likelihoods are certainly necessary, especially given the sensitivity of inland subglacial system to basal traction loss from meltwater injection \citep{schoof_ice-sheet_2010, meierbachtol_basal_2013, dow_upper_2014, doyle_persistent_2014, poinar_limits_2015}.

%%%%%%%%%%%%%%%%%%%%%%%%%%%%%%%%%%%%%%%%%%%%%%%%%%%
\section*{Acknowledgements}
We thank O. Gagliardini (University of Grenoble Alpes) for the helpful advice on Elmer/Ice simulations and N.B. Coffey (Stanford University) for the insightful discussions throughout the development of this work.
%%%%%%%%%%%%%%%%%%%%%%%%%%%%%%%%%%%%%%%%%%%%%%%%%%%
\section*{Funding statement}
The authors acknowledge partial support from NSF’s Office of Polar Programs through OPP-2344690.
%%%%%%%%%%%%%%%%%%%%%%%%%%%%%%%%%%%%%%%%%%%%%%%%%%%
\section*{Declaration of interests} 
Competing interests: The authors report no conflicts of interest.
%%%%%%%%%%%%%%%%%%%%%%%%%%%%%%%%%%%%%%%%%%%%%%%%%%%
\section*{Author ORCIDs}
Joshua H. Rines: https://orcid.org/0000-0002-3743-1096\\
Ching-Yao Lai: https://orcid.org/0000-0002-6552-7546\\
Yongji Wang: https://orcid.org/0000-0002-3987-9038\\
%%%%%%%%%%%%%%%%%%%%%%%%%%%%%%%%%%%%%%%%%%%%%%%%%%%
\section*{Author contributions}
Conceptualization: J.H.R. and C.-Y. L.; Formal Analysis: J.H.R., Y. W., C.-Y. L.; Funding acquisition: C.-Y. L.; Writing – original draft: J.H.R.; Writing – review and editing: J.H.R., Y. W., C.-Y. L.
%%%%%%%%%%%%%%%%%%%%%%%%%%%%%%%%%%%%%%%%%%%%%%%%%%%

%%%%%%%%%%%%%%%%%%%%%%%%%%%%%%%%%%%%%%%%%%%%%%%%%%%
%%%%%%%%%%%%%%%%%%% APPENDIX %%%%%%%%%%%%%%%%%%%%%%
%%%%%%%%%%%%%%%%%%%%%%%%%%%%%%%%%%%%%%%%%%%%%%%%%%%
\appendix
%%%%%%%%%%%%%%%%%%%%%%%%%%%%%%%%%%%%%%%%%%%%%%%%%%%
\section{Elmer boundary conditions and SIA details}\label{sec:appendixA_bcs_and_uL}
In the Elmer/Ice simulations, we set the inflow/outflow boundary conditions to be consistent with the sliding-accommodated shallow-ice approximation (SIA), prescribing a vertical profile of horizontal velocity given as
\begin{subequations}\label{eqn:SIA}
\begin{align}
    u(z) &= \frac{2A}{n+1}\,(\rho g \alpha)^n\Bigl[\,h^{n+1}-(h-z)^{n+1}\Bigr] + \left( \frac{\rho g h \alpha}{C_b} \right)^{1/m} \, ,
\end{align}
\end{subequations}
which represent the depth-dependent boundary conditions at the left and right sides of the Elmer/Ice simulations.  To obtain values for the far-field depth-averaged velocities, we depth-average equation \ref{eqn:SIA}:
\begin{subequations}
\begin{align}
    \bar{u} = u_L \equiv \frac{2A}{n+2}(\rho g \alpha)^n h^{n+1} + \left( \frac{\rho g h \alpha}{C_b} \right)^{1/m} 
\end{align}
\end{subequations}
This is the background velocity, $u_L$, which are subtracted from the full velocity to characterize the perturbed velocity, $u_p$ (e.g., equation \ref{eqn:up_sliding}).  Similarly, background surface velocities from SIA are given by:
\begin{subequations}
\begin{align}\label{eqn:SIA_sliding_surf}
    u\big|_h = \frac{2A}{n+1}\,(\rho g \alpha)^nh^{n+1} + \left( \frac{\rho g h \alpha}{C_b} \right)^{1/m} \, .
\end{align}
\end{subequations}

%%%%%%%%%%%%%%%%%%%%%%%%%%%%%%%%%%%%%%%%%%%%%%%%%%%
\section{Coupling length versus $\lambda$}\label{sec:appendixB_patchlen_choice}
\begin{figure}
    \centering
    \includegraphics[width=1.0\linewidth]{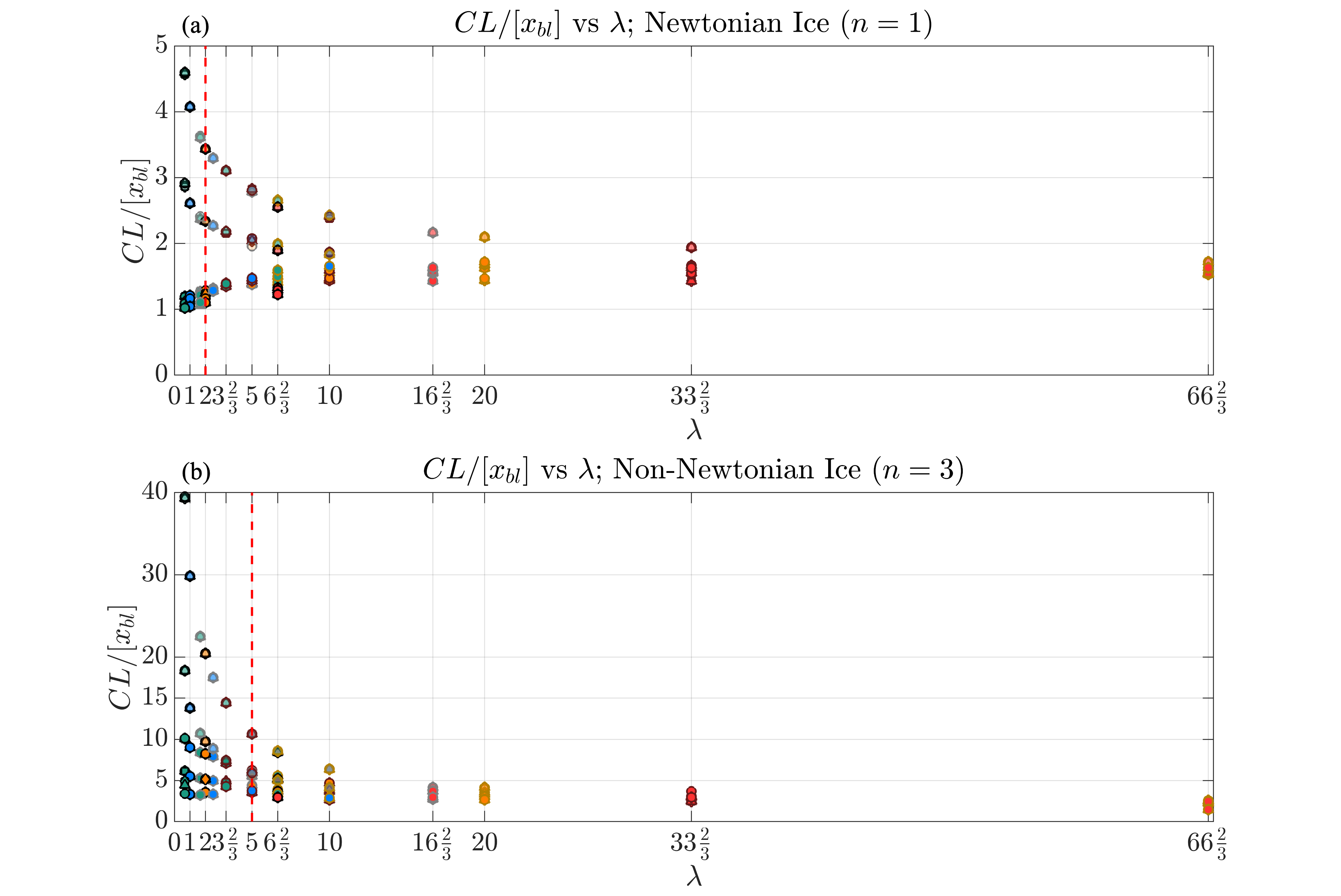}
    \caption{Numerically-computed coupling length as a function of patch length-to-thickness ratio, $\lambda$ for Newtonian (a) and non-Newtonian (b) experiments.  The vertical red dashed lines indicate the cutoff values used in our analysis in the main text.}
    \label{fig:labmda_justification}
\end{figure}

Our results are built upon the framework set up in the analytical model, which implicitly assumes that a free slip solution exists in the space within the patch bounds.  In practice, we would expect that this might not be the case for smaller patches, where more complicated dynamics are sure to be at play as the flow at the upstream and downstream edges of the patch are likely to be aware of each other.  For the analytical cases of $n=m=1$ the coupling length scales with ice thickness, and as such we might expect to see, as the slippery patch length shrinks to approach the ice thickness, that our approach becomes less defensible.  Indeed, as the ratio of patch length to thickness, $\lambda$, approaches 1, the coupling length calculated from the numerical experiments diverges from the theoretical coupling length scaling (figure \ref{fig:labmda_justification}a).  In the more general scenarios (e.g., for $n=3$), this effect is even more pronounced (figure \ref{fig:labmda_justification}b).  We thus elect to limit our analysis in the above work to $\lambda\geq2$ for Newtonian ice cases and $\lambda\geq5$ for non-Newtonian ice cases.

%%%%%%%%%%%%%%%%%%%%%%%%%%%%%%%%%%%%%%%%%%%%%%%%%%
\section{Boundary layer analysis, the case of sliding}\label{sec:appendixC_asymptotics}
In the case where sliding is allowed in the regions outside the slippery patch, a more precise boundary analysis approach may be taken.  In particular, we here carry out this procedure to solve for the stress upstream from the patch onset, taking the value of $\rho g \alpha \ell/4$ as a right-side boundary condition for the problem (e.g., equation \ref{eqn:BC_txx_transpt}).  The model is thus as follows:
\begin{equation}
    2h(\tau_{xx})_{xx} - CA\tau_{xx} = 0 \, ,
\end{equation}
which we rescale by $\tilde{\tau}_{xx} = \tau_{xx} / \rho g h \alpha$ and $\tilde{x}=x/\ell$ to give the dimensionless model (dropping tildes immediately):
\begin{equation}
    2\lambda^{-2}(\tau_{xx})_{xx} - \gamma \tau = 0 \, ,
\end{equation}
where $\lambda = \ell/h$ and $\gamma$ is defined in equation \ref{eqn:gamma} as the ratio of shear to sliding velocities, which we take to be an $\mathcal{O}(1)$ quantity here.  We can see that for cases where $\lambda>1$, the factor in front of the first term, $(\tau_{xx})_{xx}$ is small.  We define this small parameter as $\epsilon\equiv\lambda^{-2}$ and rewrite accordingly:
\begin{equation}
    \epsilon(\tau_{xx})_{xx} - \frac{\gamma}{2}\tau_{xx} = 0 \, .
\end{equation}
The leading-order outer solution is then, as we take $\epsilon\to0$, simply
\begin{equation}\label{eqn:leading_outer_soln}
    \tau_{xx}(x) = 0 \, .
\end{equation}
This is consistent as we move toward the left of the domain, toward $x\to-\infty$.  But on the right side of the domain, we impose
\begin{equation}
    \tau_{xx}\left(x=-\frac{1}{2}\right) = \frac{\lambda}{4} \,,
\end{equation}
which can not be satisfied by the leading order outer solution, $\tau_{xx}(x) = 0$ (equation \ref{eqn:leading_outer_soln}), unless we have no slippery patch in which case $\lambda=0$ gives a trivial solution.  To resolve this, we rescale the region close to the patch onset (located at $x=-1/2$) as
\begin{equation}
    \xi = \frac{x+\frac{1}{2}}{\epsilon^{\beta}} \, ,
\end{equation}
where $\beta=1/2$ is required to maintain leading order balance of both terms within the boundary layer.  Our rescaled equation is
\begin{equation}
    (\tau_{\xi\xi})_{\xi\xi} - \frac{\gamma}{2}\tau_{\xi\xi} = 0 \,,
\end{equation}
the general solution to which is
\begin{equation}
    (\tau_{\xi\xi})_{\xi\xi} = A\exp\left({\sqrt{\frac{\gamma}{2}}\xi}\right) + B\exp\left({-\sqrt{\frac{\gamma}{2}}\xi}\right).
\end{equation}
Enforcing matching with the outer solution to the left, we have
\begin{equation}
    \lim_{\xi\to-\infty}\tau(\xi) = \lim_{x\to-1/2}\tau(x) = 0 \, ,
\end{equation}
which fixes $B=0$.  The boundary condition on the right sets $A=\lambda/4$.  Composing the composite solution, we obtain:
\begin{equation}
    \tau_{xx}(x) = \frac{\lambda}{4}\exp{\left( \sqrt{\frac{\gamma}{2}}\left( \frac{x+\frac{1}{2}}{\epsilon^{1/2}} \right) \right)} \, ,
\end{equation}
or
\begin{equation}
    \tau_{xx}(x) = \frac{\lambda}{4}\exp{\left( \sqrt{\frac{\gamma}{2}}\lambda\left( x + \frac{1}{2} \right) \right)}.
\end{equation}
If we re-dimensionalize the solution (all variables now dimensional), we arrive at:
\begin{equation}
    \tau_{xx}(x) = \frac{\rho g \alpha \ell}{4}\exp{\left( \sqrt{\frac{\gamma}{2h^2}} \left( x+\frac{\ell}{2} \right) \right)}\, ,
\end{equation}
the same as equation \ref{eqn:soln_txx_bl_sliding}.

%%%%%%%%%%%%%%%%%%%%%%%%%%%%%%%%%%%%%%%%%%%%%%%%%%%
%%%%%%%%%%%%%%%%% REFERENCES %%%%%%%%%%%%%%%%%
%%%%%%%%%%%%%%%%%%%%%%%%%%%%%%%%%%%%%%%%%%%%%%%%%%%
\bibliographystyle{jfm}
\bibliography{paper_JFM}

\end{document}